\documentclass[draft]{agujournal2019}
\usepackage{url} 
\usepackage{lineno}
\usepackage[inline]{trackchanges} 
\usepackage{soul}
\usepackage{amsmath,amssymb,amsfonts}

\draftfalse

\journalname{Journal of Advances in Modeling Earth Systems (JAMES)}

\begin{document}

%
%


\title{Bringing statistics to storylines: rare event sampling for sudden, transient extreme events}

%
%




\authors{Justin Finkel\affil{1}, Paul A. O'Gorman\affil{1}}


\affiliation{1}{Department of Earth, Atmospheric, and Planetary Sciences, Massachusetts Institute of Technology}




\correspondingauthor{Justin Finkel}{ju26596@mit.edu}

\begin{keypoints}
\item Rare event algorithms may help address the challenge of simulating extreme weather events and quantifying their probability.
\item When the event of interest is sudden and transient, perturbed ensembles diversify too slowly for standard rare event algorithms to work.
\item Using the Lorenz-96 model as a prototype for midlatitude weather, we use early perturbation and a rejection step to gain a speedup.  
\end{keypoints}

\begin{abstract}
    A leading goal for climate science and weather risk management is to accurately model both the physics and statistics of extreme events. These two goals are fundamentally at odds: the higher a computational model's resolution, the more expensive are the ensembles needed to capture accurate statistics in the tail of the distribution. Here, we focus on events that are localized in space and time, such as heavy precipitation events, which can start suddenly and decay rapidly. We advance a method for sampling such events more efficiently than straightforward climate model simulation. Our method combines elements of two recent approaches: 
    adaptive multilevel splitting (AMS), a rare event algorithm that generates rigorous statistics at reduced cost, but that does not work well for sudden, transient extreme events; and ``ensemble boosting'' which generates physically plausible storylines of these events but not their statistics. We modify AMS by splitting trajectories well in advance of the event's onset following the approach of ensemble boosting, and this is shown to be critical for amplifying and diversifying simulated events in tests with the Lorenz-96 model. Early splitting requires a rejection step that reduces efficiency, but nevertheless we demonstrate improved sampling of extreme local events by a factor of order 10 relative to direct sampling in Lorenz-96. Our work makes progress on the challenge posed by fast dynamical timescales for rare event sampling, and it draws connections with existing methods in reliability engineering which, we believe, can be further exploited for weather risk assessment. 
\end{abstract}

\section*{Plain Language Summary}
What is the strongest rainstorm that we can expect in a given thousand-year period? To augment the available $\sim100$ years of historical data and to account for climate change, computer simulations are a useful, but expensive, tool to answer such questions. A model must run for many millennia to deliver an answer with statistical confidence. \emph{Rare event algorithms} provide a promising alternative simulation protocol, in which an ensemble of short simulations is biased to produce more extreme events and reweighting is used to correct for the bias when calculating statistics. However, a classical rare event algorithm fails when the events of interest are short and ``bursty'' (like heavy rainstorms) instead of long and slow-moving (like anomalously hot summers). We modify the rare event algorithm to make it amenable to precipitation-like events in an idealized dynamical system with chaotic traveling waves. 

%
%

%


%
%
%
%

\section{Introduction}

In climate modeling, high spatial resolution is important for realistically representing localized extreme weather events like cyclones producing extreme precipitation and winds \cite{OBrien2016resolution,vanderWiel2016resolution}. But given finite computational resources, high resolution has to be traded off with the need for ensembles of models and simulations to deal with uncertainty related to model physics, parameters, initial conditions and boundary conditions including emissions scenarios. Extreme events are particularly challenging because they occur infrequently, and hence need large ensemble sizes to have their small probabilities accurately quantified. The conflict for computational resources therefore comes to a head in the study of extreme events. 

A variety of shortcuts have developed in the past century to alleviate this conflict.  Leading statistical approaches include extreme value theory \cite<EVT;>[]{Coles2001introduction} and large deviation theory \cite{Touchette2009large}, which respectively describe the behavior of \emph{maxima} and anomalously large \emph{running means} in random processes. In principle, we can use these theories to fit a parametric family to limited data and then extrapolate to even longer return periods. EVT has become an important tool in risk assessment and climate change attribution \cite{Kharin2007changes,Naveau2020statistical}, while large deviation theory succinctly encodes the severity of long-lasting, large-area events such as persistent heat waves \cite{Galfi2021applications}.   Statistical theories help make the most of a fixed dataset, but parameter estimation can be unstable given the restrictive underlying assumptions and the limited datasets available \cite{Huang2016estimating,Galfi2017convergence}. For example, EVT only holds in the limit of large blocks of data or high thresholds for extremity, which directly conflicts with the requirement of many samples for low-variance parameter estimation. Moreover, statistical theories don't provide spatio-temporal resolved extreme events (e.g., the spatial field of rainfall and other fields  on the day of an extreme event) which are needed to drive impact models.

Statistical or dynamical downscaling is another way to address the problem of extremes by reducing the computational cost of obtaining high-resolution output from long simulations or large ensembles \cite{Huang2020future,Lee2020statistical,Emanuel2021response,Saha2022downscaling,Krouma2022assessment}. 
Downscaling nevertheless has some drawbacks.  Dynamical downscaling using regional climate models faces the challenge of correctly forcing a regional model with output from a different global model, and the regional model inherits errors in large-scale fields from the global model \cite{Adachi2020methodology}, while   statistical downscaling assumptions can create systematic errors \cite{Schmidli2007statistical} and may not generalize to different climates.   

The focus of this paper is \emph{rare event sampling}, which is a strategy for allocating more of the computational effort towards rare events, and less effort towards the long intervening periods of comparatively mild behavior. This is usually achieved by \emph{splitting} methods, which consist of three steps repeated in a cycle: (1) run an ensemble of simulations forward, (2) identify the ensemble members making the most progress towards the extreme event, and (3) clone these most-promising ensemble members (applying small perturbations) while discarding the less-promising members, resulting in a new ensemble that is more prone to extremes than was the original ensemble. With repeated rounds of splitting, one can populate the tail of the probability distribution more fully, while neglecting the more typical behavior of lesser interest. Crucially, in statistical analysis of the ensemble, one must compensate for the bias by weighting each clone with a factor less than one, relying on the \emph{importance sampling} formalism. See \citeA{Bucklew2004introduction} for an introduction to rare event sampling. 

This generic procedure has many possible variants, which have been developed largely in the fields of physics \cite{Kahn1951estimation,Giardina2006direct}, chemistry \cite{Kastner2011umbrella,Zuckerman2017weighted}, and reliability engineering \cite{Au2001estimation}, but have recently started to make an impact on Earth and planetary sciences. For example, extreme European heat waves were sampled by \citeA{Ragone2018computation} and \citeA{Ragone2021rare} with genealogical particle analysis (GPA), and by \citeA{Yiou2020simulation} with empirical importance sampling. \citeA{Wouters2023rare} sampled extreme European seasonal precipitation accumulations, also using GPA. \citeA{Webber2019practical} developed a quantile-based variant of GPA to sample more extreme versions of tropical cyclones. Planetary science applications include jet nucleation  \cite{Bouchet2019rare2} and orbit destabilization \cite{Abbot2021rare}. For studies of climate, rare event sampling can be applied to global models or paired with the dynamical and statistical downscaling approaches mentioned earlier.

We have elected to use a particular rare event algorithm called \emph{adaptive multilevel splitting} (AMS), which was first established by \citeA{Cerou2007adaptive} and is similar to the earlier RESTART algorithm  \cite{Villen1991restart}. \citeA{Lestang2018computing} successfully applied AMS to the Ornstein-Ulhenbeck process, while \citeA{Lucente2022coupling} and \citeA{Baars2021application} used AMS to study regime transitions in idealized climate models. AMS has also been usefully employed in other diverse fields such as molecular dynamics and air traffic control (see \citeA{Cerou2019adaptive} for a recent review). The distinguishing feature of AMS is that it operates on the level of full trajectories over a fixed time horizon, and applies the small perturbation to trajectories at the instant that they first cross a threshold of extremity. The ``child'' trajectory is identical to its parent up until this time, whereas it diverges from its parent afterward to give a new realization of the extreme event. All ensemble members failing to cross the threshold are discarded, and the threshold is then raised for repeated rounds of splitting and killing. 

A related approach, ``ensemble boosting'', is a novel technique for generating ``storylines'' of unprecedented climate extremes \cite{Gessner2021very,Gessner2022physical}. In this approach, one identifies several extreme events from a long climate simulation, perturbs the antecedent conditions (1-3 weeks ahead of time), and re-simulates the event to generate alternative realities, which sometimes turn out even more extreme. While similar to splitting methods, ensemble boosting does not explicitly quantify statistics. As explained below, a major goal of this paper is to combine the benefits of ensemble boosting with that of rare event algorithms, in particular AMS. 

Given the successes in using rare event sampling discussed above, it is desirable to also use it to sample shorter-term extreme weather events, such as daily precipitation extremes, which have large societal impacts in the current climate \cite{Wright2021resilience,Thompson2017unseen} and are expected to intensify under climate change \cite{OGorman2015precipitation,Pfahl2017understanding,Tandon2018understanding,Myhre2019frequency}. However, heavy precipitation events (or high wind events) have some dynamical characteristics that distinguish them from the previous applications and pose challenges to existing rare event algorithms. Unlike continental-scale, seasonally averaged  anomalies studied previously \cite{Ragone2018computation,Wouters2023rare}, heavy precipitation events of interest are often sudden, transient, and relatively small-scale. Their timescale at a particular location is often limited by the propagation of the dynamical feature causing the precipitation such as cyclones and fronts \cite{Dwyer2017changing}. The strategy used in \citeA{Ragone2018computation} and \citeA{Wouters2023rare} relies on some slow-moving notion of \emph{progress} towards the extreme event, naturally given by the integrated temperature anomaly itself when targeting extreme seasonal average temperatures, in order to decide which simulations to clone or kill. 
In the precipitation study of \citeA{Wouters2023rare}, the extreme event is again a seasonal total, for which a mid-seasonal total is a reasonable measure of progress.  But for individual precipitation events, if one uses precipitation itself to measure progress towards the event, and applies perturbations to a simulation when precipitation picks up, it is too late for these perturbations to take effect by the time of maximum precipitation. The event simply comes and goes faster than perturbed simulations diverge. \citeA{Lestang2018computing} found a similar pathology with AMS when sampling extreme pressure fluctuations on a body embedded in a turbulent channel flow. There, the extreme events were caused by vortices sweeping past the body, roughly analogous to cyclones sweeping past a location on Earth, and the rapidity of the fluctuation crippled the effectiveness of the standard splitting strategy. 

\begin{figure}
    \centering
    \includegraphics[width=0.99\linewidth,trim={0 9cm 6cm 0},clip]{"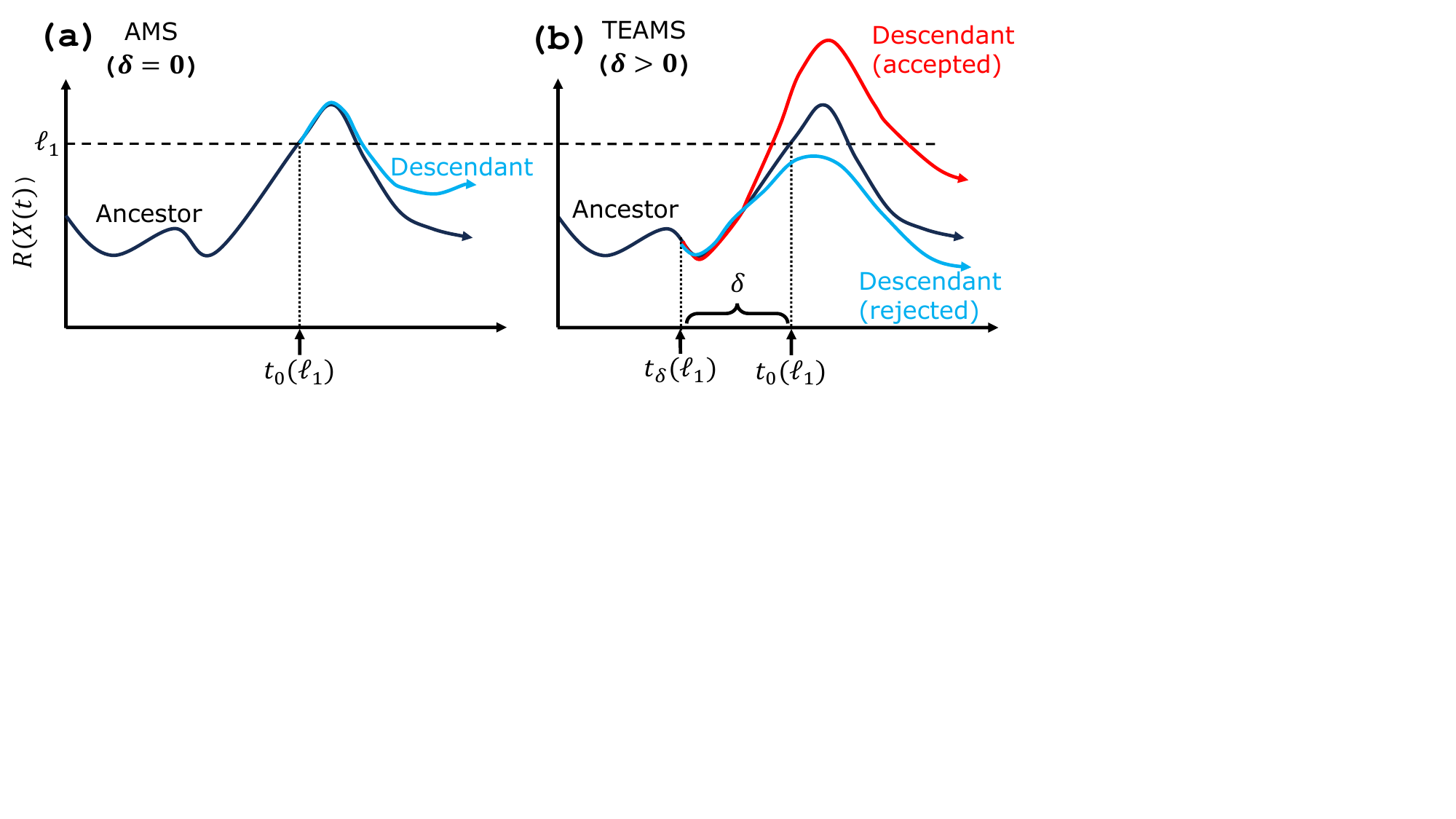"}
    \caption{Schematic of the splitting step in (a) AMS and (b) TEAMS. Black curves represent an initial ensemble member, or ancestor, which exceeds the first level $\ell_1$ and has been selected for cloning in the first round. In AMS, the perturbation is applied at the instant $t_0(\ell_1)$ when the ancestor first exceeds $\ell_1$, resulting in a descendant trajectory (blue) which essentially replicates the extreme event because the separation timescale is longer than the event itself. On the other hand, in TEAMS (right) we apply the perturbation in advance, by some margin $\delta>0$. This can sometimes result in rejection (blue descendant), i.e., failure to cross $\ell_1$. However, when a descendant is accepted (red) it will be more distinct from the ancestor than the corresponding descendant in AMS and have the potential to reach a substantially higher peak value.}
    
    \label{fig:teams_schematic}
\end{figure}

To isolate and solve the problem of applying rare event algorithms to sudden, transient extremes, we postpone the specific application to precipitation and first descend the model hierarchy to the Lorenz-96 model \cite{Lorenz1996predictability}, a spatiotemporal chaotic system often used as a toy model for the atmosphere. The model produces extreme events posing the same algorithmic challenges as precipitation extremes: intermittent, short-lived bursts carried by traveling waves with unpredictable amplitudes. It has been used in numerous past studies of extreme event statistics and predictability \cite{Sterk2017predictability,Qi2016predicting,Hu2019effects}. With this cheap but behaviorally rich model, we have developed a simple modification to AMS, drawing inspiration from ensemble boosting by simply applying a split in advance of the event's onset by some advance split time $\delta$---hence, ``trying early'' AMS (TEAMS). To make this statistically rigorous, a rejection step is necessary, which comes at an efficiency cost, but still enables moderate speedups of $\sim$10 relative to direct sampling. Fig.~\ref{fig:teams_schematic} displays a schematic diagram for TEAMS, which will be elaborated in section~\ref{sec:subset}. In fact, TEAMS is a repurposing of a more general method called \emph{subset simulation} \cite{Au2001estimation} from structural reliability engineering, a field whose sophisticated rare event algorithms could benefit the climate risk community.

This paper is organized as follows. In section~\ref{sec:L96}, we present a stochastically forced Lorenz-96 model and the behavior of its extreme events as a function of stochastic forcing strength. In section~\ref{sec:subset},
we first introduce the general framework of subset simulation. In section~\ref{sec:ams}, we specialize to AMS, and in section~\ref{sec:ams_failure} we show that AMS fails in the low-noise forcing regime, which is often most relevant for weather and climate models. In section~\ref{sec:trying_early}, we modify AMS to use a ``trying early'' step with rejection sampling and recover a substantial speedup. In section~\ref{sec:choosing_delta}, we further explore the relationship between the advance splitting time---a key algorithmic parameter---and classical notions of predictability timescales. Finally, in section~\ref{sec:conclusion} we point out directions for further development.

\section{Lorenz-96: a customizable spatiotemporal chaotic system}
\label{sec:L96}

\begin{table}
    \caption{Physical parameters for Lorenz-96 system (upper section), and algorithmic parameters for the TEAMS algorithm (lower section).}
    \centering
    \begin{tabular}{c|l|l}
    \hline
    Symbol & Explanation & Value or range \\
    \hline
    $K$ & Number of longitude sites & 40 \\
    $a$ & Strength of advection term & $\{1, 0\}$ (mostly 1) \\
    $F_0$ & Constant background forcing & 6 \\
    $m$ & Wavenumber for stochastic forcing & $\{1,4,7,10\}$ (mostly 4) \\
    $F_m$ & Strength of stochastic forcing at wavenumber $m$ & $\{3, 1, 0.5, 0.25, 0 \}$ \\ 
    \hline
    $N$ & Number of initial ensemble members & 128 \\
    $\kappa$ & Number of members to kill each round & 1 \\
    $J$ & Number of rounds of splitting & 896 \\
    $T$ & Time horizon & 6 \\
    $\delta$ & Advance split time & $[0,2]$ \\
    \hline
    \end{tabular}
    \label{tab:parameters}
\end{table}

\citeA{Lorenz1996predictability} introduced a simple dynamical system (L96 hereafter) meant to capture some crucial aspects of atmospheric dynamics. 
The model state space consists of $K$ ($\geq4$) variables $\{x_k\}_{k=1}^K$ arranged on a one-dimensional periodic lattice, each $k$ representing a longitude sector on Earth. 
$x_k$ represents a generic atmospheric variable like wind speed or vorticity and evolves according to the coupled equations
\begin{equation}
    \label{eq:L96}
    \frac{dx_k}{dt}=ax_{k-1}(x_{k+1}-x_{k-2})-x_k+\mathcal{F}_k,\hspace{1cm}k=0,...,K-1,
\end{equation}
where $x_{k+K}$ is identified with $x_k$. The quadratic terms on the right-hand side represent advection, like the quadratic nonlinearity in the material derivative of the Navier-Stokes equations, which on its own conserves ``energy'' $\frac12\sum_kx_k^2$. The linear term $-x_k$ represents damping due to friction, and the additive term $\mathcal{F}_k$ represents external forcing, like a meridional insolation gradient. The latter two terms destroy exact energy conservation, but balance out in a time-averaged sense to make for a statistically steady state. \citeA{Lorenz1996predictability} introduced the above model with $\mathcal{F}_k$ constant in $k$ and also a version in which $\mathcal{F}_k$ is a ``subgrid-scale forcing'' that is a function of an additional tier of dynamical variables representing finer scales, and this version has proven useful for testing stochastic parameterization schemes 
\cite<e.g.,>[]{Wilks2005effects,Hu2019effects,Gagne2020machine}.
Here, we also allow $\mathcal{F}_k$ to vary stochastically with longitude ($k$) and time: 
\begin{equation}
    \label{eq:noise}
    \mathcal{F}_k=F_0+F_m\bigg[\eta_1\cos\bigg(\frac{2\pi mk}{K}\bigg)+\eta_2\sin\bigg(\frac{2\pi mk}{K}\bigg)\bigg]
\end{equation}
where $\eta_{1,2}$ are independent Gaussian white-noise processes, and $m$ is an integer wavenumber. Formally, Eq.~\eqref{eq:noise} renders Eq.~\eqref{eq:L96} a diffusion process, using the It\^{o} convention for stochastic integrals \cite{Pavliotis2014stochastic}. This simple stochastic forcing is analagous to a stochastic parameterization in a weather or climate model, and in the AMS framework it allows us to easily generate new ensemble members by splitting an existing ensemble member at a certain time. We verify below that for weak amplitudes the stochastic forcing does not appreciably alter model statistics. 

The parameters used here are summarized in the upper section of Table~\ref{tab:parameters}. We set $K=40$, following \citeA{Lorenz1998optimal}. We fix the constant part of the forcing to be $F_0=6.0$, which is sufficient for weak turbulence (a larger value would be needed with smaller $K$).  We choose the stochastic forcing wavenumber as $m=4$ because that empirically seems to drive ensemble members apart slightly faster than very small or large wavenumbers (see section \ref{sec:prt}). Indeed  the stochastically perturbed parameterization tendencies (SPPT) method developed at ECMWF uses noise that is spatially correlated at a $\sim10^\circ$ length scale   \cite{Buizza1999stochastic,Palmer2009stochastic}.  The amplitude of $F_m(=F_4)$ will be explored systematically below.  One further parameter, the coefficient $a$, determines the strength of the advection term. $a=1$ is standard for L96, while $a=0$ gives an array of correlated Ornstein-Uhlenbeck (OU) processes \cite{Pavliotis2014stochastic}. Retaining the OU process as a special case of L96 is useful to  provide a reference case on which existing rare event splitting algorithms excel. Results for $a=0$ are shown in supplementary Figs.~S1 and S2, and all other results presented are for $a=1$. 

\begin{figure}
    \centering
    \includegraphics[width=0.99\linewidth,trim={0 0cm 12cm 0},clip]{"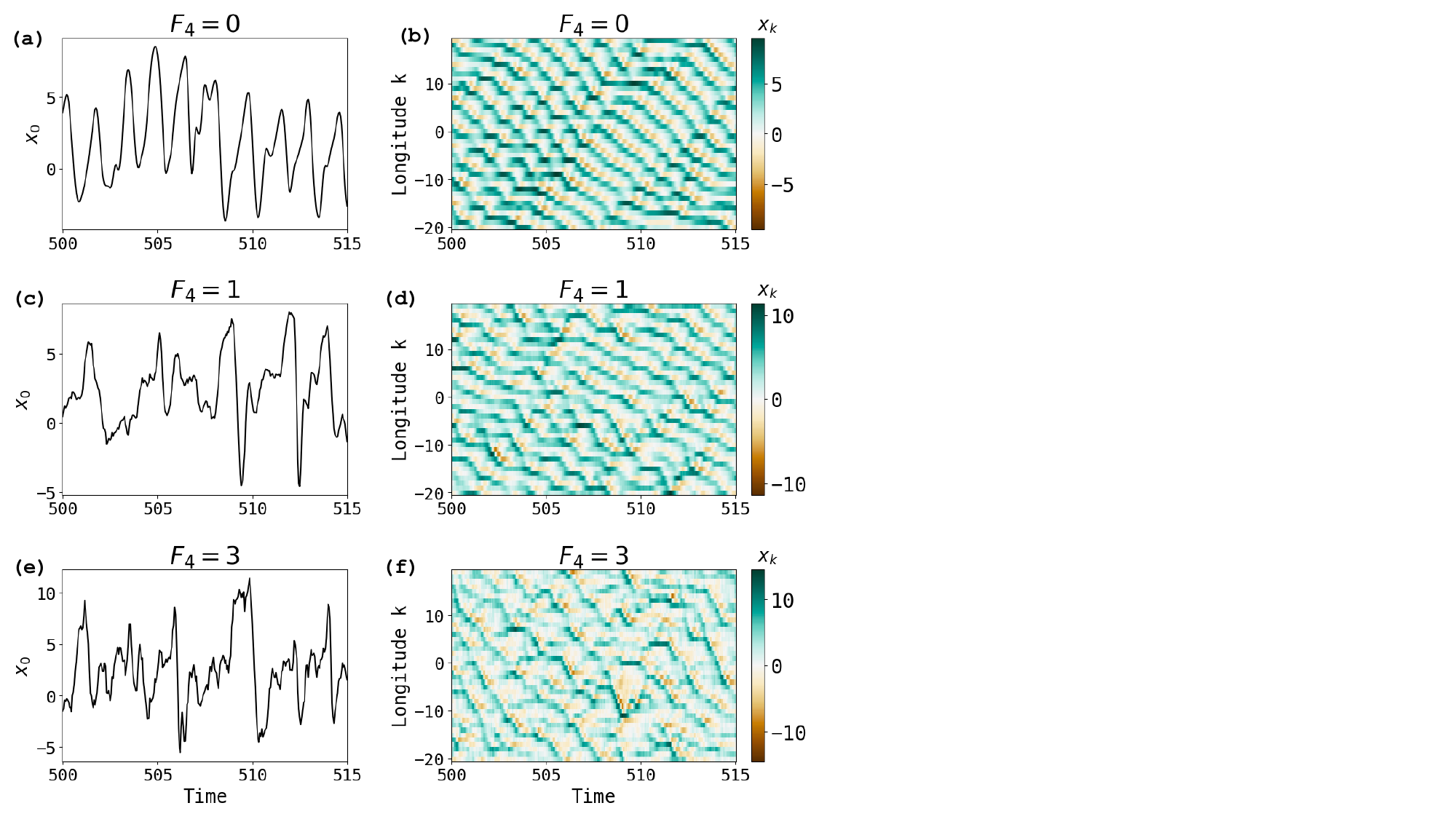"}
    \caption{Time evolution of the L96 model expressed as timeseries of $x_0(t)$ (left column) and Hovm\"oller diagrams (right column) with three different levels of stochastic forcing. (a,b) have $F_4=0$ (the deterministic system); (c,d) have $F_4=1$ (moderate forcing); (e,f) have $F_4=3$ (strong forcing).} 
    \label{fig:dns_integrations}
\end{figure}

Fig.~\ref{fig:dns_integrations} displays short numerical integrations of L96 with four different parameter choices. We used the Euler-Maruyama method with a timestep of 0.001 to integrate Eq.~\eqref{eq:L96}, saving out every 0.05 time units. For comparison, \citeA{Lorenz1998optimal} interpret a single time unit as 5 days. The left column shows single-site variables $x_0(t)$ for each parameter set, while the right column shows corresponding Hovm\"oller diagrams. In the standard deterministic system $F_4=0$ in the top row, $x_0(t)$ fluctuates with a semi-regular period of $\sim2$ time units (10 ``days'') but with irregular amplitudes, the largest of which are precisely the extreme events we choose to study here. The Hovm\"oller diagram reveals these fluctuations to arise from a field of traveling waves, with roughly eight peaks and troughs moving with negative (``westward'') phase velocity. 
The waves experience intermittent disturbances, sometimes getting stuck in place for several turnover times and setting up favorable conditions for extreme events. Globally, these stagnations manifest as kinks that propagate in the positive (``eastward'') direction. This is reminiscent of atmospheric Rossby waves, whose phase and group velocities have opposite signs (up to a Doppler shift due to the mean flow) \cite{Lorenz1998optimal}. 
Thus, we can loosely think of the peaks and troughs as being like highs and lows in the midlatitude atmosphere.


Fig.~\ref{fig:dns_integrations} rows 2 and 3 show analogous pictures for moderate ($F_4=1$) and strong $(F_4=3)$ stochastic forcing, respectively. As noise increases the traveling waves transition from unidirectional to zigzagging. The timeseries become more jagged and more liable to take large excursions from their mean and hover there for longer durations. 

\begin{figure}
    \centering
    \includegraphics[width=0.99\linewidth,trim={0 8cm 0cm 0},clip]{"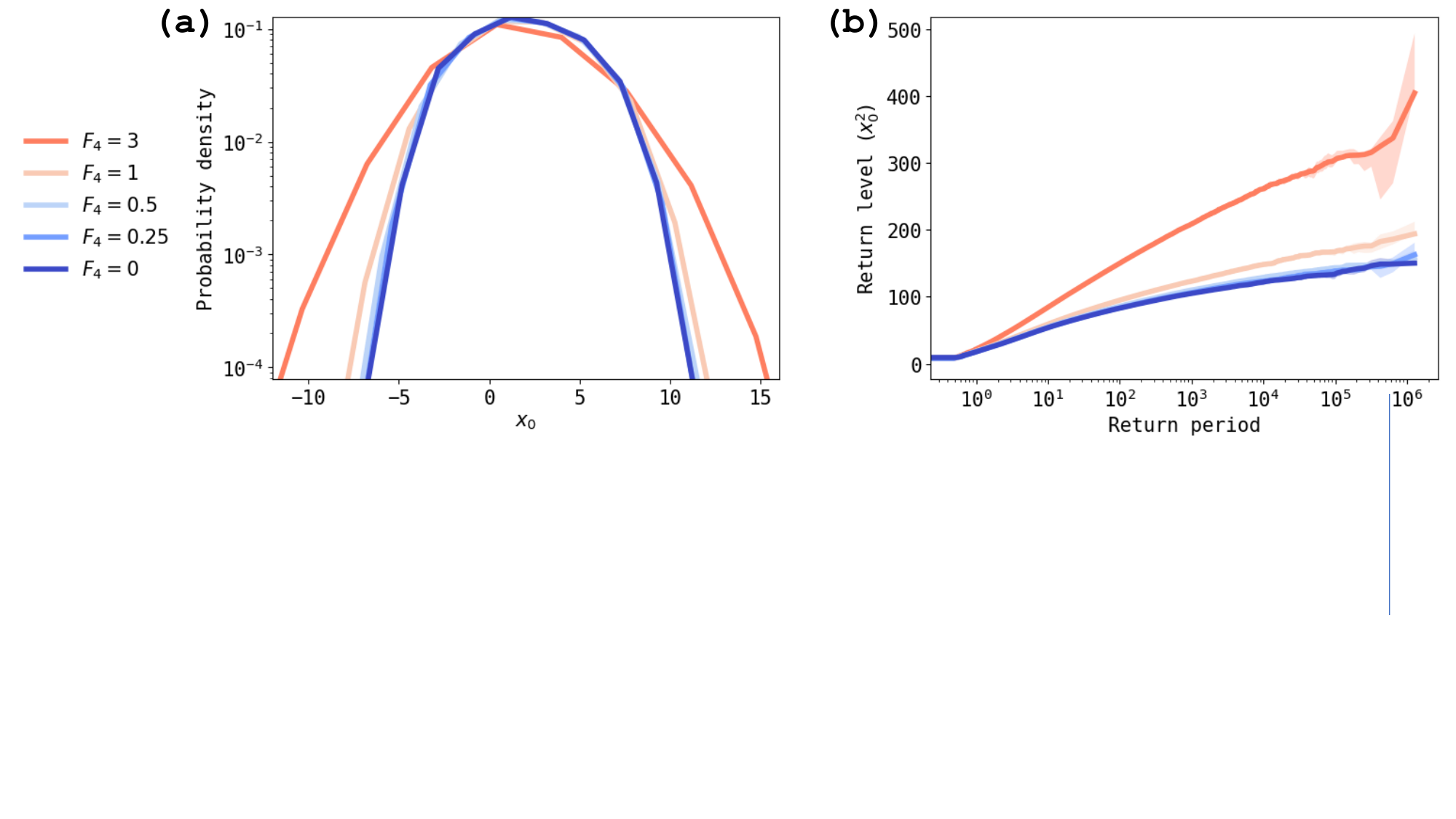"}
    \caption{Steady-state statistics of the L96 model as a function of noise strength, calculated from a long simulation of length $1.28\times10^6$. (a) Histograms of the model variable at one site ($x_0$) and (b) return level vs. return period for (twice) the local energy  $x_0^2$. Shading in (b) represents 95\% bootstrapped confidence intervals from the modified block maximum method. See text for details. }
    \label{fig:dns_pdfs}
\end{figure}

Fig.~\ref{fig:dns_pdfs}a overlays PDFs of the single-site value ($x_0$) for all these parameter regimes, plus two more: $F_4=0.5$ and $0.25$. Reducing the noise roughly preserves the mode but shrinks the tails. The PDF appears basically converged for $F_4\leq0.5$. Fig.~\ref{fig:dns_pdfs}b confirms this is true even in the far tail, with a log-transformed plot of return level vs. return time for $x_0^2$. The limiting case $F_4=0$ has a bounded tail, which is easy to see with an energy argument (see also \citeA{Qi2016predicting}): defining $\overline{x}=\frac1K\sum_{k=1}^Kx_k$, the energy $E=\frac12\sum_kx_k^2$ evolves as $\frac{dE}{dt}=-2E+KF\overline{x}$.  Since $|\overline{x}|\leq\sqrt{\overline{x^2}}=\sqrt{2E/K}$ by the Cauchy-Schwarz inequality, the first term dominates for $E$ larger than some critical $E_0$, which must therefore bound the steady-state distribution's tail. However, $E_0$ would increase with $K$, i.e., higher-dimensional systems can in principle support heavier tails \cite<e.g.>[ch. 4 discusses general relationships between the shape parameter and the attractor dimension]{Lucarini2016extremes}. This is part of our motivation to set $K$ relatively large.

The return level vs. return period plot (as in Fig.~\ref{fig:dns_pdfs}b) will be used throughout the paper, and we calculate it using the ``modified block maximum'' method of \citeA{Lestang2018computing}. For a fixed \textit{return level} $\ell$, the \textit{return period} $\tau(\ell)$ is defined as the mean (over initial conditions and noise realizations) of the waiting time until an exceedance occurs: $\tau(\ell)=\mathbb{E}[\min\{t:R(x(t))>\ell\}$], where $R$ is some observable of interest for the dynamical system. We take $R(x)=x_0^2$, the local energy (times two) at longitude $k=0$. 
\citeA{Lestang2018computing} approximates the exceedance times by a Poisson process for high $\ell$ to give
\begin{equation}
    \label{eq:mbm}
    \tau(\ell)=-\frac{T}{\log\big[1-p_T(\ell)\big]}.
\end{equation}
where $p_T(\ell)$ is the probability of at least one exceedance in a fixed time $T$. $p_T(\ell)$ can be estimated from any collection of length-$T$ blocks of data---\emph{either from a single continuous timeseries or not}. This is very useful because rare event splitting algorithms generate branching trees of short trajectories, from which we can estimate block-wise exceedances but not return times directly.

To produce Fig.~\ref{fig:dns_pdfs}b, we started with simulations of length $1.28\times10^6$ (after discarding the first 50 for spinup), split them into $B$ blocks of length $T=6$, and measure the maxima $M_1,...,M_B$ of $x_0^2$ over each block. Letting $M_{(b)}$ denote the $b$th largest block maximum, we use the empirical (complementary) CDF estimator, $\widehat{p}_T(M_{(b)})=b/B$. Hence, the return curve should interpolate the ordered pairs $(\tau_b,\ell_b)=\big(-\frac{T}{\log(1-b/B)},M_b\big)$.  Because it is common to think of $\ell$ as a function of $\tau$, and to consider logarithmically spaced return periods, we linearly interpolate $M_{(b)}$ over $\log\tau_B$ to get a curve $\hat{\ell}(\tau)$.  We bootstrap to estimate uncertainty, resampling the blocks $1,...,B$ with replacement and repeating the above procedure 5000 times.  Shading indicates the basic bootstrap 95\% confidence interval \cite{Wasserman2004all}, meaning $\hat{\ell}(\tau)+(\hat{\ell}(\tau)-\ell^*_{0.975}(\tau),\hat{\ell}(\tau)-\ell^*_{0.025}(\tau))$, where $\ell^*_\alpha$ denotes the $\alpha$th quantile of the bootstrap distribution of $\hat{\ell}$ for each $\tau$. Note that when $\ell^*_{0.025}(\tau)$ is much less than $\hat{\ell}(\tau)$, we get a very large \emph{upper} bound on the confidence interval, because it suggests via the basic bootstrap philosophy that $\hat{\ell}(\tau)$ could be very much less than the true parameter $\ell(\tau)$. 
The lowest-noise curves are close to within uncertainty even in the far tails, demonstrating the convergence of extreme value statistics for $F_4\leq0.5$. This confirms that stochastic forcing, when sufficiently weak, does not alter the system's statistics very much, which allows us to approximate the deterministic system's rare events while remaining within the AMS framework which relies on explicit randomness.

The longest return period estimable by this method of ``direct numerical simulation'' (DNS) is $\sim8\times10^5$, the simulation's length. Rare event algorithms can sample physical realizations of extreme events at long return periods $\tau(\ell)$ with much less computation time than $\tau(\ell)$, but have not yet been applied to local events in L96 with weak stochastic forcing. 
\citeA{Wouters2016rare} did apply rare event algorithms to L96, but their system parameters differed substantially from ours, with $F_0=256$ giving a much more turbulent regime reminiscent of a stochastic process. Moreover, their target quantity of interest was a globally averaged energy, whereas we target local energy at one longitude as a closer analogue to extreme precipitation or winds hitting a particular location.

The parameters $a$ and $F_4$ allow us to test the performance of AMS for a range of problems, from systems on which AMS performs well to more difficult systems akin to the extreme local precipitation problem. $a=0$ (the OU process) is an easy setting for AMS; $a=1$ with large noise $F_4$ is harder, but still doable because of the dominance of noise. Shrinking $F_4$ further, towards the system of actual interest, gradually renders standard AMS ineffective and leads us to a modified version of the algorithm called TEAMS that allows for early splitting. The next sections present the basic algorithm and its modification along this parameter path. 

\section{Subset simulation}
\label{sec:subset}

TEAMS (and the special case AMS) may be viewed as a version of \emph{subset simulation} (SS), which we use to frame our overall approach, and which we believe has considerable potential for application to climate problems. SS was introduced in \citeA{Au2001estimation} and has been most widely used in structural reliability engineering \cite{Huang2016assessing}. For a short pedagogical introduction, see \citeA{Zuev2015subset}. The description below will introduce several tunable algorithmic parameters, which are summarized in the lower section of Table~\ref{tab:parameters}.

The goal is to estimate the probability that a random variable $x$ from a distribution $\rho$ gives rise to large values of some quantity of interest $S(x)$,
\begin{equation}
    p(\ell)=\int\mathbb{I}\{S(x)>\ell\}\rho(x)\,dx=\mathbb{E}_\rho\big[\mathbb{I}\{S(X)>\ell\}\big],
\end{equation} 
given only the ability to draw samples $X_1,X_2,...\sim\rho$. $\mathbb{I}\{\cdot\}$ denotes the indicator function: one if the argument is true, zero if false. For us, each $X_i=\{X_i(t):0\leq t\leq T\}$ is a length-$T$ trajectories of L96 (with stochastic forcing); the score function is a maximum over the interval, $S(X)=\max_{0\leq t<T}R(X(t))$; and $\rho(x)$ is the distribution over \emph{trajectories of length} $T$ induced by the stochastically forced L96 system. In structural engineering, $X$ might be the state of a building or dam, with $\rho(x)$ induced by a probability distribution over external stresses like wind, earthquakes, or rainfall, while $S(x)$ would measure the proximity to failure. Because the probabilities of interest are very small, a set of independent samples $\{X_n\}_{n=1}^N$ from $\rho$ will usually have few if any exceedances, making the ``vanilla'' Monte Carlo estimate of $p(\ell)$ (the fraction of exceedances) subject to high relative uncertainty. The ratio of the estimator's variance to its mean is approximately $1/\sqrt{Np(\ell)}$ \cite{Zuev2015subset}. If we want to aim for a tenfold-longer return period with the same uncertainty, we need to generate tenfold more samples. Worse, to reduce uncertainty tenfold we would need one hundredfold more samples, which may be untenable. 

SS breaks down this task into a sequence of easier tasks by setting up a series of intermediate levels $\ell_1<\ell_2<...<\ell_J=\ell$ where $J$ is the number of levels, and estimating a sequence of conditional probabilities $\mathbb{P}\{S(X)>\ell_{j+1}|S(X)>\ell_j\}=:p(\ell_{j+1}|\ell_j)$, which all have moderate magnitudes and are expected to be easier to estimate. Their product provides an estimate for the target probability:
\begin{equation}
    \hat{p}_{\mathrm{SS}}(\ell)=\hat{p}(\ell_1)\hat{p}(\ell_2|\ell_1)...\hat p(\ell_J|\ell_{J-1}).
\end{equation}
The first term can be estimated by vanilla Monte Carlo: generate $N$ samples $X_1,...,X_N$, and attach unit weights to each: $W_n=1$ for $n=1,...,N$. Rank the samples by $S$ so that $S(X_{(1)})\leq S(X_{(2)})\leq...\leq S(X_{(N)})$, and let $\hat p(\ell_1)=(N-\kappa_1)/N$, where $\kappa_1$ is chosen so that $S(X_{(\kappa_1)})\leq\ell_1<S(X_{(\kappa_1+1)})$.  The parameter $\kappa_1$ is the number of trajectories that are ``killed'' meaning they don't appear in the first subset (see below).  For the case of AMS, $\kappa_1$ is chosen as a parameter of the algorithm,  and $\ell_1$ is then set adaptively as $\ell_1=\frac12[S(X_{(\kappa_1)})+S(X_{(\kappa_1+1)})]$.

The second term $\hat p(\ell_2|\ell_1)$ is estimated with a splitting strategy  in which we focus in on the ``subset'' of samples that exceed the first threshold: $\{S(X)>\ell_1\}$ containing samples $X_{(i)}$ with $\kappa_1< i\leq N$. To better sample this subset, we spawn additional samples from it via a ``Modified Metropolis algorithm'':
\begin{enumerate}
    \item Initialize a list $\mathbb{X}_1=\{X_{(\kappa_1+1)},...,X_{(N)}\}$, which will eventually grow to a (user-chosen) size $N_1$ as well as a first-in-first-out queue $\mathbb{Q}$ of the same elements but in a random order: the ``parent queue''.
    \item Pop $\mathbb{Q}$ to yield the next parent $X$. Apply some small perturbation to $X$ to generate a new sample $\widetilde{X}$, which itself is drawn from $\rho$ but correlated to $X$.  A general way to do this is with one step of the Metropolis-Hastings algorithm which involves an accept/reject step, but an easier approach is available in the particular case of AMS as described in the next section.
    \item Evaluate $S(\widetilde{X})$. If it exceeds $\ell_1$, we have successfully generated a new sample from the subset. Accept the new sample, meaning insert $\widetilde{X}$ into both $\mathbb{Q}$ and $\mathbb{X}_1$ and assign it a weight equal to that of its parent  $X$. Otherwise, if $S(\widetilde{X})\leq\ell_1$, reject $\widetilde{X}$. Re-insert $X$ into $\mathbb{Q}$ and add a copy of $X$ to $\mathbb{X}_1$. (In implementation, we don't store two copies of the high-dimensional object $X$, but rather we assign a multiplicity to each member and increment $X$'s multiplicity by one.) 
    \item Repeat steps 2 and 3 until $\mathbb{X}_1$ has $N_1$ elements (counting multiplicity). 
    \item Multiply the weights of all members of $\mathbb{X}_1$ by a factor $(N-\kappa_1)/N_1$, which preserves the total weight $N$ of the original ensemble $\{X_n\}_{n=1}^N$ while spreading that weight over more members.
\end{enumerate}

Having expanded to $N_1$ samples from the subset $\{S(X)>\ell_1\}$, we can now proceed to the next level and generate additional samples from the next subset $\{S(X)>\ell_2\}$ so that it contains $N_2$ samples, where $\ell_2$ can be determined adaptively as an order statistic of $\mathbb{X}_1$, i.e., the average of the $\kappa_2$th and the $(\kappa_2+1)$th ranked values. The same procedure is repeated to generate the next subset $\mathbb{X}_2$ (and $\mathbb{Q}$ is initialized with only unique elements, not counting multiplicity, in order to maintain as much diversity as possible). $\mathbb{X}_3,\mathbb{X}_4,...,\mathbb{X}_J$ are generated in the same fashion, until either a computational budget is reached, an ultimate target threshold is overcome, or some other halting criterion is met. 

Ultimately we are left with a weighted ensemble $\{(X_1,W_1),...,(X_M,W_M)\}$, where $M=\kappa_1+\kappa_2+...+\kappa_J+N_J$. The sampling $\{S(X_m)\}_{m=1}^M$ is over-represented in the tails, but with correspondingly smaller weights there, and all weights sum to $N$. Any expectation of an observable $\Phi(x)$ can be estimated as 
\begin{equation}
    \mathbb{E}[\Phi(X)]=\int\Phi(x)\rho(x)\,dx\approx\hat{\Phi}=\frac1N\sum_{m=1}^M\Phi(X_m)W_m.
\end{equation}
The SS algorithm will generally help to improve this estimate for functions $\Phi$ most sensitive to the tail region of $S(x)$, rather than its central bulk. In particular, setting $\Phi(x)=\mathbb{I}\{S(x)>\ell\}$, we recover the estimator $\hat{p}_{\mathrm{SS}}(\ell)$:
\begin{align}
    \mathbb{E}[\mathbb{I}\{S(X)>\ell\}]=p(\ell)\approx\frac1N\sum_{m:S(X_m)>\ell}W_m=\hat{p}_{\mathrm{SS}}(\ell).
\end{align}

An important set of algorithmic choices are the population parameters $N,N_1,...,N_J$, the killing numbers $\kappa_1,\kappa_2,...,\kappa_J$, as well as the halting criterion which determines $J$. \citeA{Cerou2019adaptive} reviews theoretical bases for several different choices, but here for simplicity we opt for the same rule as used in \citeA{Lestang2018computing}: $\kappa_j=\kappa=1$ (the ``drop 1'' rule) and $N_j=N$ for all $j=1,...,J$ (the population is replenished after each new level is set). Note that with $\kappa_j=1$, only a single parent is selected from $\mathbb{Q}$ at each round before the level is raised and the queue re-initialized. 

\subsection{Adaptive multilevel splitting (AMS)}
\label{sec:ams}

AMS (in particular ``trajectory AMS (TAMS)'' in the nomenclature of \citeA{Lestang2018computing}) can be seen as a special case of SS where each $X=\{X(t):0\leq t\leq T\}$ is a length-$T$ trajectory of a stochastic dynamical system, $S(X)=\max_{0\leq t<T}R(X(t))$ for some time-dependent score function $R$, and with a particular choice for splitting trajectories.
 Trajectories are split by constructing a new forcing sequence $\widetilde{\eta}(t)$ ($\widetilde{\eta}_{1,2}(t)$ for our L96 model) to drive the child trajectory $\widetilde{X}(t)$ starting from the old forcing sequence $\eta(t)$ that drove the parent. First, copy the initial condition $\widetilde{X}(0)=X(0)$. Then, copy $\widetilde{\eta}(t)=\eta(t)$ up until some \emph{split time} $t_{\mathrm{sp}}$, which is chosen as first time $t_0(\ell)$ that the parent clears the threshold: 
\begin{equation}
    t_{\mathrm{sp}}=t_0(\ell_1)=\min\{t\in[0,T]:R(X(t))>\ell_1\}.
\end{equation}
For following times $t\geq t_{\mathrm{sp}}$, swap in a new and independent noise forcing sequence for $\widetilde{\eta}(t)$. No Metropolis-style accept/reject step is needed for step (2) above; each newly sampled Brownian increment of $\widetilde{\eta}(t)$ is drawn independently from $\mathcal{N}(0,\Delta t)$, and so $\widetilde{\eta}(t)$ is a proper sample from the same noise-generating distribution as $\eta(t)$. Furthermore, the choice of $t_{\mathrm{sp}}=t_0(\ell_1)$ guarantees $\widetilde{X}(t)=X(t)$ for all $t\leq t_0(\ell_1)$, so that $S(\widetilde{X})>\ell_1$, and acceptance is guaranteed in step (3) as well. 

The change in forcing for $t\geq t_{\mathrm{sp}}$ will cause the child to diverge from the parent, producing a new---but correlated---sample (Fig.~\ref{fig:teams_schematic}a). How correlated $\widetilde{X}$ is to its parent $X$ depends on $t_{\mathrm{sp}}$, with later $t_{\mathrm{sp}}$ implying a longer shared history and less independence. Applying the split at $t_{\mathrm{sp}}=t_0(\ell)$ maximizes the independence of the child---and ultimately the diversity of the AMS ensemble---while guaranteeing $S(\widetilde{X})$ exceeds $\ell_1$, and therefore is accepted in the modified Metropolis Algorithm. The same procedure is carried out for every subsequent level. 

We performed a sequence of AMS experiments with the following parameters:
\begin{enumerate}
	\item Physical constants and timescales:  $F_4\in\{3,1,0.5,0.25\}$ for the default case $a=1$  which gives the stochastically forced L96 model, and $F_4=3$ for the case $a=0$ which gives the OU process (shown in supplementary Figs.~S1 and S2).
 We fix $F_0=6$, and $K=40$ throughout, and set the time horizon to $T=6$.
	\item Ensemble sizes and population control: $N=N_j=128$ and $\kappa_j=1$ for $j=1,2,...,J=896$ adhering to a fixed computational budget of 1024 time horizons simulated. One additional halting criterion is imposed: if the population loses so much diversity that all active ensemble members descend from the same ancestor, we terminate the algorithm early. 
        \item  We repeat the whole procedure $M=56$ times for each parameter set, with different seeds for pseudo-random number generation.  Each repetition will be called a ``run'' of AMS. Having multiple runs allows us to assess variance, and by using pooled estimates from all runs to hedge against stagnation within local optima of phase space in a particular run.
\end{enumerate}

The initial $N$-member ensemble is generated as a sequence of consecutive blocks from a moderate initialization simulation of length $N\times T$ ($T=6$ is the time horizon), after discarding the first 50 units as spinup. The spinup is initialized as $x_k(0)=F_0+\frac{1}{1000}\sin\big(\frac{2\pi k}{K})$. The random number generator used to create the noise forcing sequences $\eta_{1,2}(t)$ is seeded with $s\in\{0,...,M-1\}$, a different value for each AMS run with a fixed parameter set. 
The $N$ initial blocks, although weakly correlated, comprise a sample from the steady-state distribution of the stochastic L96 system. 
Larger $N$ reduces the variability of the AMS results, but it also means more up-front cost and more rounds of splitting needed to reach return times long enough to make the algorithm worthwhile. 

We compare our results from AMS to a long DNS simulation of length $1.28\times10^6$ (separate from the initialization), which is then further elongated by a factor of 40 (concatenating all $K$ timeseries end-to-end) into $5.12\times10^7$, exploiting the statistical equivalence of all $K=40$ sites of L96. This curve is our best estimate of ground truth. Note that the symmetry is only exploited to extend the DNS estimate, not the AMS estimate. In a climate model with zonal inhomogeneities, such as continents, it would be inappropriate to aggregate different longitudes together. 

\begin{figure}
    \centering
    \includegraphics[width=0.7\linewidth,trim={0 0cm 21cm 0},clip]{"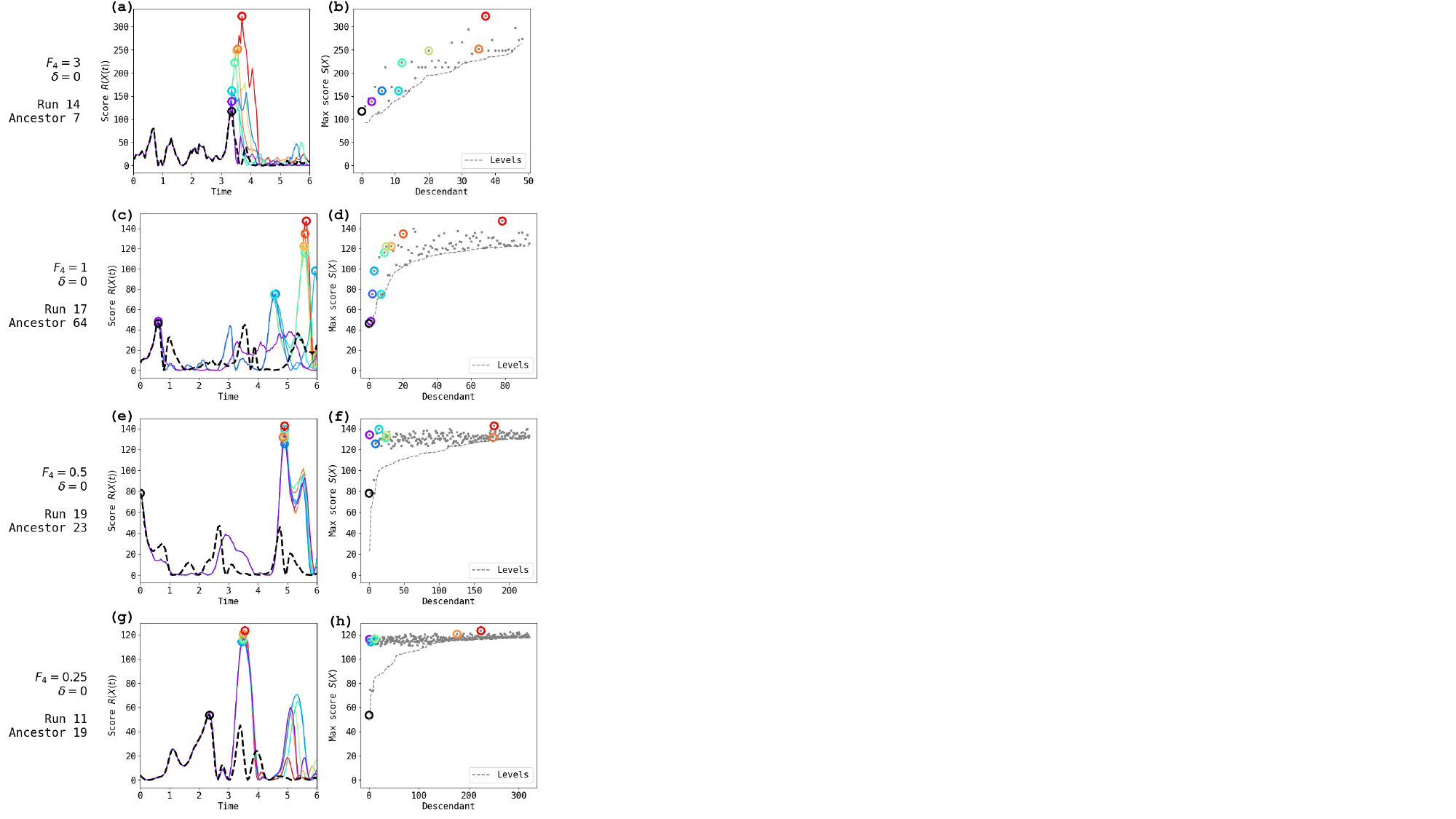"}
    \caption{
    Scores for single ancestors and their descendents within the AMS algorithm (special case of TEAMS with $\delta=0$). For each stochastic forcing amplitude, 56 independent runs of AMS were carried out (indexed 0-55) with $N=128$ ensemble members (0-127). (a) Time-dependent score function $R(X(t))$ for the 7th initial ensemble member (ancestor) of run 14 for $F_4=3$. A black circle indicates the scalar score $S(X)=\max_tR(X(t))$. $R(X(t)$ and $S(X)$ are also shown for a single lineage (path down the family tree) in a sequence of brightening colors, ending with the highest scoring descendant's score in red. (b) Scores in gray dots, with the horizontal axis numbering all descendants from ancestor 7 of run 14 for $F_4=3$. Colored circles indicate those descendants in the lineage from (a). The dashed gray curve indicates the levels $\ell$ from which each descendant was split. (c,e,g) are the same as (a), and (d,f,h) are the same as (b), but with stochastic forcing strength decreasing to $F_4=1,0.5$, and 0.25 respectively. In each case, the run and ancestor were hand-selected among the ancestors with the maximum boosting.}
    \label{fig:storylines_ams}
\end{figure}

\begin{figure}
    \centering
    \includegraphics[width=0.99\linewidth,trim={0 0 15cm 0},clip]{"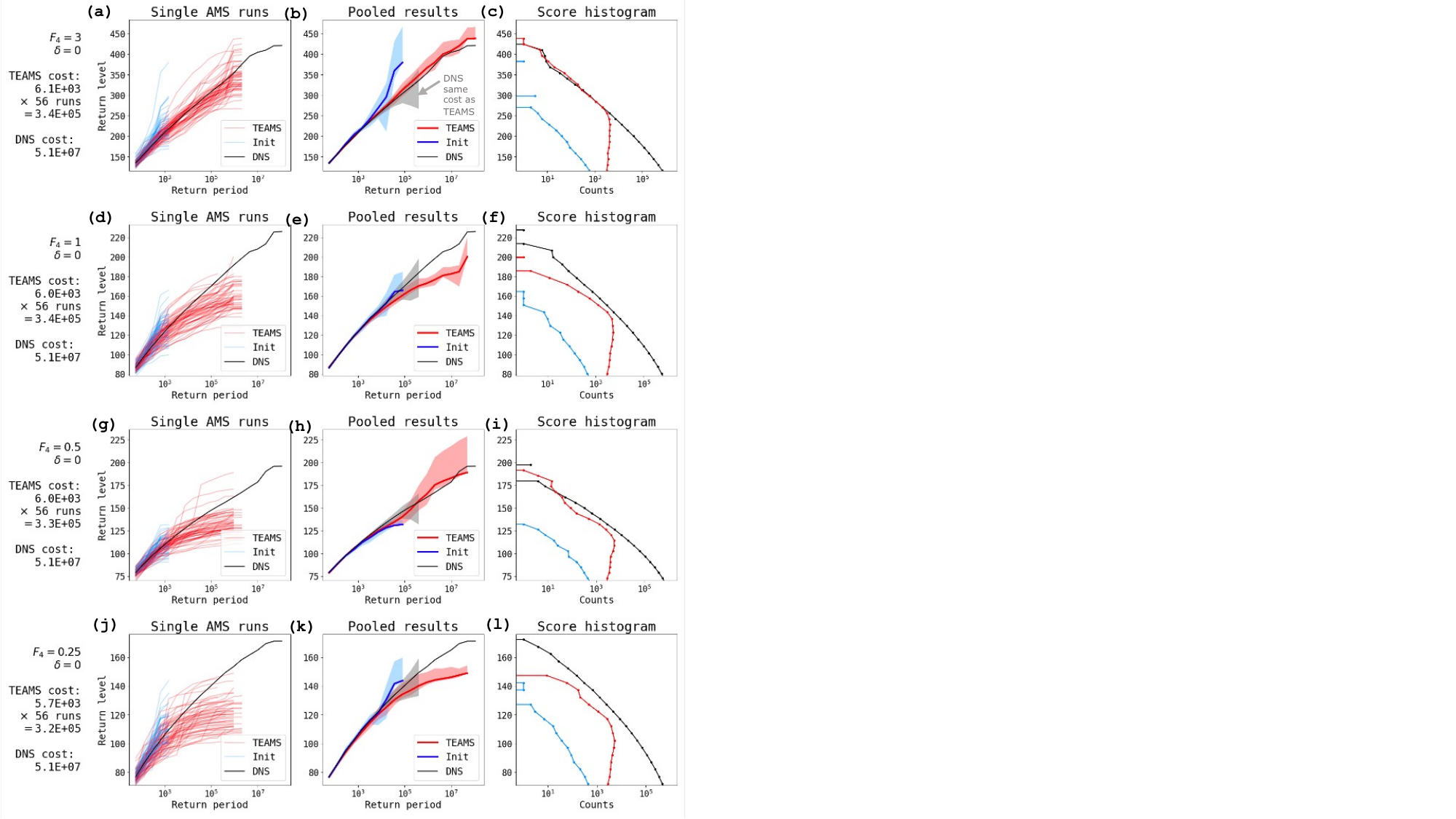"}
    \caption{Performance of the AMS algorithm (special case of TEAMS with $\delta=0$).
    (a) Return level vs. return period plots for $F_4=3$. Blue lines show estimates from the initial 128 members of each AMS run; red lines show estimates from the completed AMS runs;  black line shows DNS.
    (b) Return level vs. return period for a pooled AMS ensemble containing all $56\times1024$ members. Blue and red envelopes indicate 95\% confidence intervals (see text for details). Gray envelope is a 95\% confidence interval based on subsets of DNS equal in total cost to the 56 AMS runs. Thus, the dashed red line and shading from AMS is of equal cost to the gray shading from DNS.
    (c) Unweighted histogram of scores for AMS initialization (blue), completed AMS (red), and DNS (black). 
    Following rows are same as first row, but with noise decreasing to $F_4=1,0.5$, and 0.25, respectively. The slight variability in TEAMS costs listed to the left are due to the early halting criterion of one single ancestor remaining (see section~\ref{sec:subset}).} 
    \label{fig:returnplots_ams}
\end{figure}

Fig.~\ref{fig:storylines_ams}a,b illustrates the effect of successive mutations over the course of the AMS algorithm, on the relatively easy test case with strong stochastic forcing, $F_4=3$ and $a=1$ (the even easier case of $a=0$---the OU process with no interference from advection---is documented in \citeA{Lestang2018computing} and included in supplementary Figs. S1 and S2 for completeness). By design, the levels increase monotonically over the course of generations and the descendant scores march upward, ultimately mutating the moderate ancestor into an extreme descendant. Going beyond this successful ``anecdote'', Fig.~\ref{fig:returnplots_ams}(a,b,c) confirm the benefit of AMS for a \emph{statistically accurate} sampling of the distribution's tails. Fig.~\ref{fig:returnplots_ams}a shows return period curves calculated with the modified block maximum method according to three datasets: the full weighted ensemble from AMS; 
the initialization (``Init''), consisting of $N$ ensemble members per AMS run; and the long DNS simulation.
The return levels are interpolated onto a common logarithmically spaced grid of return periods for easy comparison between the three data sources. Whereas return level estimates based on the initializations alone (blue) scatter considerably around the ground truth, AMS provides a tighter range of estimates (red) around the ground truth, and for $\sim3$ orders of magnitude-longer return periods, at only 8 times the cost of initialization (1024 members from an initial 128). Moreover, each AMS run is $\sim5000$ times less costly than the DNS run that gave the ground truth curve; altogether, the 56 AMS runs are $\sim100$ times less costly.


Another way of comparing AMS to DNS is by pooling together all members from the 56 ensembles and considering them as one larger ensemble of size $56\times1024=57344$. Fig.~\ref{fig:returnplots_ams}b shows the resulting statistics which have the advantage of extending to considerably longer return periods than the individual AMS runs. Here, as in Fig.~\ref{fig:dns_pdfs}, the error bars are given by the basic bootstrap 95\% confidence interval using 5000 bootstrap samples, but in the case of DNS (gray error bar), each bootstrap resampling contains only enough blocks to match the total simulation time used by AMS (including all independent runs). This lets us compare the uncertainties fairly between the two methods. In the case of AMS error bars, the members within a single run are not independent of each other, and so we resample the AMS runs. That is, we sample the numbers $\{0,...,55\}$ 5000 times with replacement, and for each resampling we pool together all members from the corresponding list of AMS runs, including repetitions. Fig.~\ref{fig:returnplots_ams}c shows the unweighted histogram of scores coming from the three data sources. The difference in shape of the AMS histogram compared to the DNS histogram demonstrates the main effect of AMS: to undersample the low end of the distribution and oversample the tail, shifting the computational burden to where it is more useful for sampling extremes. 

We consider AMS to ``win'' over DNS if either of two criteria are met: (i) the AMS estimate remains close to the DNS (relative to error bar width) for return periods well beyond the AMS total simulation time $T_{\mathrm{AMS}}$; (ii) the AMS error bar is much smaller than the DNS error bar at $T_{\mathrm{AMS}}$. Under strong stochastic forcing, AMS performs very well by both criteria, accurately (and confidently) estimating return periods as long as $10^7$ in the pooled estimate using only $3.4\times10^5$ time units of computation. This aligns with the demonstration in \citeA{Lestang2018computing} for the OU process, and serves as a departure point for our modification of the algorithm.

\subsection{Failure of AMS in the regime of weak stochastic forcing}
\label{sec:ams_failure}

The story gets more complicated when the stochastic forcing is weak and nonlinear dynamics dominate. In deterministic chaos, perturbations grow exponentially with a rate inversely proportional to the \emph{Lyapunov timescale}---at least, so long as the perturbations remain infinitesimal. Only after several elapsed Lyapunov times---what we call the \emph{divergence timescale}, quantified further in section~\ref{sec:choosing_delta}---do perturbations become large enough to be useful for splitting algorithms, but also at which size nonlinear effects take over. In contrast to deterministic chaos, white noise realizations diverge immediately. The stochastic L96 system inherits both behaviors to some extent, determined by the relative strength of stochastic forcing. Our main thesis is that when nonlinear dynamics dominate, and divergence time exceeds the duration of the event of interest, standard AMS is inadequate, but this can be remedied by adjusting the choice of splitting time $t_{\mathrm{sp}}$ as shown in the next section. 

Fig.~\ref{fig:storylines_ams}c-h show ancestors and descendents for AMS, analogous to Fig.~\ref{fig:storylines_ams}a,b and with identical algorithmic parameters, but with decreasing levels of stochastic forcing: $F_4=1,0.5,0.25$.  For all four stochastic forcing strengths, ancestors can spawn more extreme descendants. However, there is a key difference between the strong- and weak-stochastic forcing regimes. With strong stochastic forcing $F_4=3$ (Fig.~\ref{fig:storylines_ams}a,b), each descendant along the lineage improves upon the \emph{same event}. In other words, the sequence of maximum scores comes from a peak in the timeseries for $R(X(t))$ that grows taller and taller, drifting only slightly forward in time. With weaker stochastic forcing (Fig.~\ref{fig:storylines_ams} c-d, e-f and especially g-h), events tend to see only modest boosts from generation to generation. The only way for a child $\widetilde{X}$ to improve \emph{substantially} over its parent $X$ is by creating a whole new event---a new peak later in the time horizon---rather than building on an existing event. This happens because the stochastic forcing is too weak to open a large gap between $R(\widetilde{X}(t))$ and $R(X(t))$ during the short interval between the splitting time $t_0(\ell)$, when $R(X(t))$ first exceeds $\ell$, and the peak $\text{argmax}_tR(X(t))$. The child ends up essentially replicating the parent's peak, which is the same behavior illustrated schematically in Fig.~\ref{fig:teams_schematic}a. The characteristic time scale of the peak (what we will call the event duration) is set by the zonal propagation of waves, and this timescale is not long enough compared to the divergence time for AMS to work well. The same phenomenon was observed in \citeA{Lestang2020numerical}): extreme spikes in the force on a body in a turbulent channel flow (see their Fig. 14) could not be boosted via AMS, which was attributed to the ``sweeping'' of vortices past the body. Similar reasoning holds for the zonal propagation of waves in L96 and the passage of midlatitude cyclones or fronts past a location in the midlatitudes.


Fig.~\ref{fig:returnplots_ams} summarizes the performance of AMS for different strengths of stochastic forcing. The suspicion of failure raised by Fig.~\ref{fig:storylines_ams} is confirmed by the clear degradation of performance as $F_4$ shrinks. In particular, the individual AMS return level curves tend to fall farther and farther underneath the true return level curves (left column of Fig.~\ref{fig:returnplots_ams}). There is a large scatter in the individual runs, and in the case $F_4=0.5$, a lucky few of the 56 runs salvage the pooled estimate for a decent approximation of the DNS return levels, but the width and asymmetry of the confidence intervals indicate the unreliability of this result (Fig.~\ref{fig:returnplots_ams}h). The problem becomes particularly acute as $F_4$ drops to 0.25, with the individual AMS runs barely improving upon the initial scores (Fig.~\ref{fig:returnplots_ams}j) and a large underestimate at longer return periods for the pooled estimate (Fig.~\ref{fig:returnplots_ams}k). 

It thus appears that standard AMS is dead on arrival for cases where the  divergence timescale is longer than the event duration. In principle, there is a canonical fix for this problem, namely to use a more intelligent score function than the quantity of interest $R(X(t))$ itself. The ideal such proxy is the \emph{committor}: the probability, given an initial condition $X(t)=x$, that $R(X(s))$ will exceed $\ell$ at some time $s\in(t,T)$ before the time horizon ends. By definition, the committor incorporates information about the model state $X(t)$ that is not available from $R(X(t))=x_0^2$, for example the speeds and magnitudes of different wave packets scattered across the domain that may all soon converge at $k=0$ and result in an extreme burst of energy. The committor is an \emph{optimal} score function for AMS in terms of minimizing the variance for $\hat{p}(\ell)$ \cite{Lestang2018computing,Cerou2019adaptive,Lucente2022coupling}. Considerable research has recently pursued approximation strategies for the committor in various climate applications \cite<e.g.,>[]{Tantet2015early,Finkel2021learning,Lucente2022committor,Miloshevich2022probabilistic,JacquesDumas2023data}. 

Unfortunately, these strategies all require either a high volume of training data---potentially canceling out the savings of a rare event algorithm, which is useful precisely in the low-data regime---or very specific knowledge of phase space geometry, such as a bistable structure, which is not typically available for realistic climate models. A second, related problem is that the optimality property only holds true for a single committor with a fixed threshold $\ell$. What if we seek return periods for a whole range of thresholds? We would have to sacrifice the accuracy of some return periods in favor of others. Alternatively, we could use the committor for a single very high threshold $\ell_{\mathrm{max}}$, but then even less training data would be available. Although it is interesting and worthwhile to search for committor functions based on traveling-wave dynamics, we leave that to future work, and in the next section we describe a simpler strategy to get around the stagnation issue seen in Fig.~\ref{fig:storylines_ams}.

\subsection{Trying-early adaptive multilevel splitting (TEAMS)}
\label{sec:trying_early}

To address the failure of AMS in the nonlinear regime, we adjust $t_{\mathrm{sp}}=t_\delta(\ell)=:t_0(\ell)-\delta$ by an \emph{advance split time} $\delta>0$, allowing some time for the child $\widetilde{X}$ to drift farther away from the parent and possibly achieve a higher maximum score. Indeed, ensemble boosting \cite{Gessner2021very} does exactly that, systematically applying perturbations every day from 19 to 7 days in advance of heat wave onset, although ensemble boosting does not by itself allow the calculation of return periods for the boosted events. When splitting early we lose the guarantee that $R(\widetilde{X}(t))$ clears the current level $\ell$ (decpicted schematically in Fig.\ref{fig:teams_schematic}b), which is why we frame our modified algorithm using subset simulation (see section \ref{sec:subset}) which includes an accept/reject step: when a child fails to score higher than $\ell$, it is discarded from the ensemble and its parent is duplicated instead (in other words, doubling its statistical weight).  The resulting algorithm, which we call TEAMS (``trying-early adaptive multilevel splitting''), incurs additional cost due to rejected samples, but also gains back the ability to build significantly upon ancestral scores. One can interpret $\delta$ as setting the width of the proposal distribution, a key parameter in Markov chain Monte Carlo methods. A wider proposal allows the child to explore farther afield from its parent, but increases the risk of rejection. Proposal width often has to be tuned carefully, and the sampling community has devoted substantial efforts to adaptively designing the proposal \cite{Gilks1998adaptive,Andrieu2008tutorial}. Such methods will surely prove useful for complex climate models, but in our present proof-of-concept study of the algorithm, we found approximately optimal $\delta$ values by exhaustive grid search for each noise level. Section \ref{sec:choosing_delta} explains this procedure and shows that the optimal $\delta$ can be related to the error saturation timescale, a classical measure of predictability.

We performed a sequence of TEAMS experiments with $(F_4,\delta)\in\{3,1,0.5,0.25\}\times\{0,0.2,0.4,...,2.0\}$. We adjust the time horizon $T=6+\delta$ to give each parameter choice the same length of score to boost. 
All other parameters are as before for the AMS experiments.

\begin{figure}
    \centering
    \includegraphics[width=0.7\linewidth,trim={0 0cm 21cm 0},clip]{"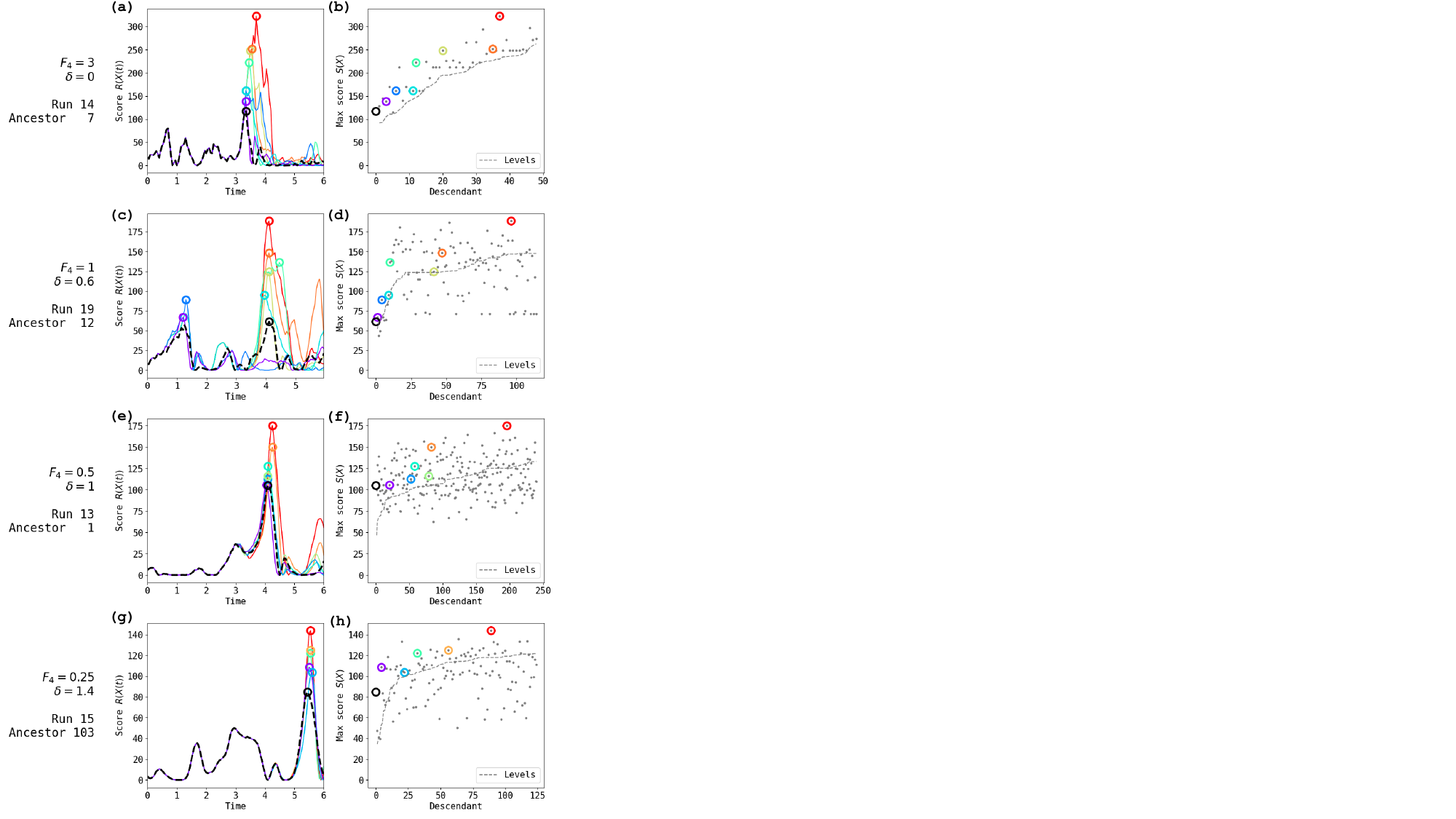"}
    \caption{Scores for single ancestors and their descendants generated by the TEAMS algorithm:  the same as Fig.~\ref{fig:storylines_ams} but with advance split times $\delta$ chosen to be approximately optimal for each noise level: $\delta=$ 0, 0.6, 1, and 1.4 for $F_4=3, 1, 0.5$, and 0.25, respectively. Because $\delta=0$ is optimal for $F_4=3$, (a,b) is the same as Fig.~\ref{fig:storylines_ams}a,b. Section~\ref{sec:choosing_delta} explains how the $\delta$ values were chosen.}
    \label{fig:storylines_teams}
\end{figure}

Fig.~\ref{fig:storylines_teams} shows TEAMS in action for the same parameter sets from Fig.~\ref{fig:storylines_ams}, but with (roughly optimal) advance splitting times $\delta=0.0,0.6,1.0$, and $1.4$ for the decreasing noise levels (at $F_4=3$, $\delta=0$ still works best, and panel (a) is the same as in Fig.~\ref{fig:storylines_ams}a)). Note that the score functions $R(X(t))$ are only defined for times $t>\delta$, because if $t_0(\ell)<\delta$ then $t_\delta(\ell)<0$, so we cannot apply the split early enough. This is implemented by setting the early scores to \texttt{NaN}, and lengthening the time horizon from $T$ to $T+\delta$ as mentioned above. We account for this extra cost in all the performance calculations to follow, but we omit the first $\delta$ time units from the plots. For all four stochastic forcing strengths, we see examples of children building significantly, and directly, upon a parent's maximum, without having to discover a new peak farther into the future. The values of the scores form continuous point clouds in panels (b,d,f,h), unlike the discrete horizontal bands appearing in Fig.~\ref{fig:storylines_ams}(f,h) where $\delta=0$ and stochastic forcing is weak. The negative side-effect is that many gray dots fall short of the gray dashed line, indicating a rejected sample. Clearly, increasing $\delta$ brings both higher risk and higher reward. 


\begin{figure}
    \centering
    \includegraphics[width=0.99\linewidth,trim={0 0 15cm 0cm},clip]{"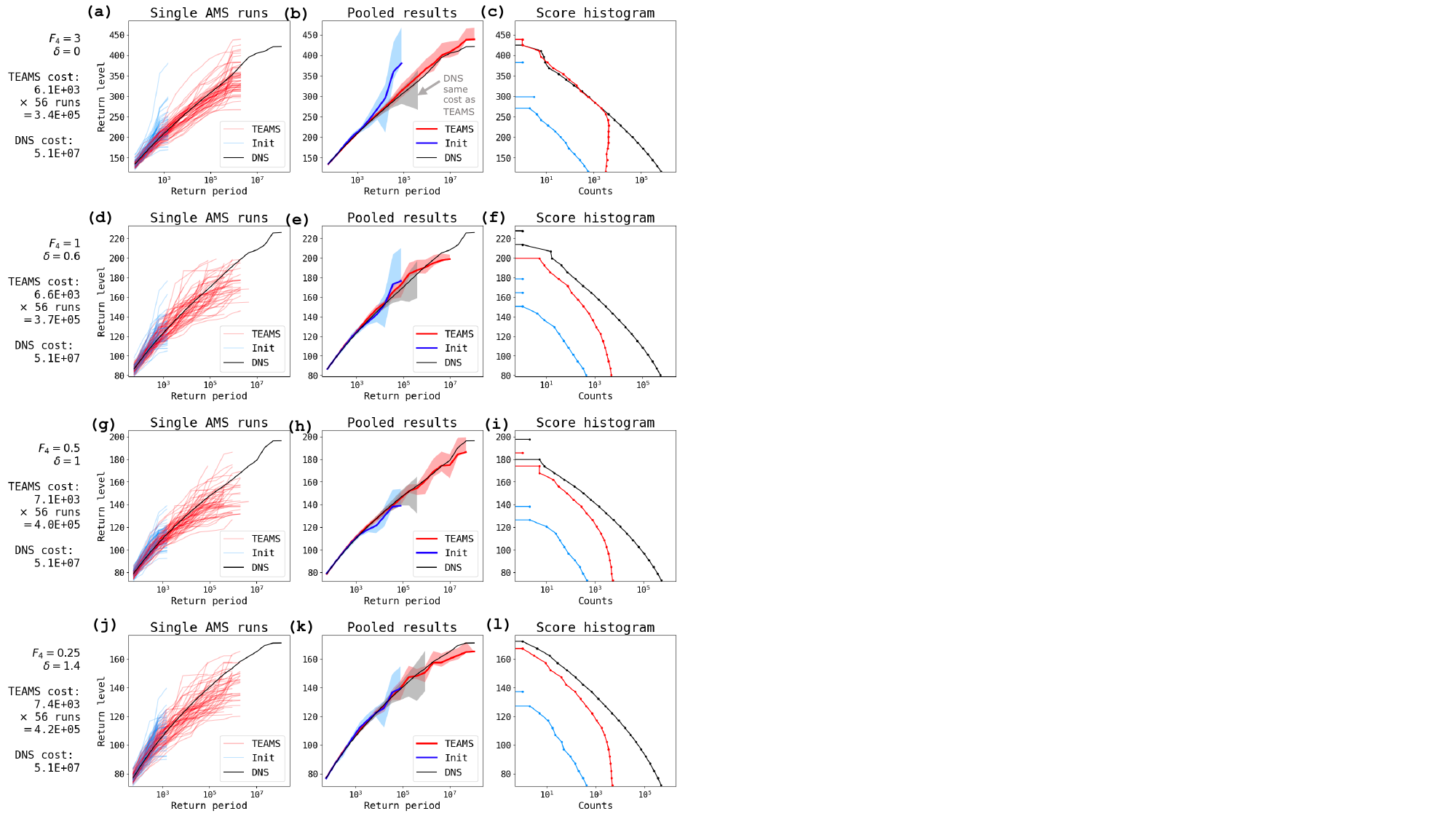"}
    \caption{ Performance of the TEAMS algorithm: the same as Fig.~\ref{fig:returnplots_ams} but with advance split times $\delta$ chosen to be approximately optimal for each noise level: $\delta=$ 0, 0.6, 1, and 1.4 for $F_4=3, 1, 0.5$, and 0.25, respectively. Because $\delta=0$ is optimal for $F_4=3$, (a-c) are the same as Fig.~\ref{fig:returnplots_ams}a-c.}
    \label{fig:returnplots_teams}
\end{figure}

Fig.~\ref{fig:returnplots_teams} quantitatively confirms the hopeful suggestion of Fig.~\ref{fig:storylines_teams}: that increasing $\delta$ can give TEAMS a speedup over DNS in the weak stochastic forcing regime. For all cases shown, TEAMS extends the estimated return period, \emph{accurately}, well beyond the gray envelope which marks the limit achievable by an equal-cost run of DNS. The black ground truth curve remains within the 95\% confidence band of TEAMS to return periods of $\sim10^7$ across all forcing levels. Simultaneously, the TEAMS confidence band is narrower than the DNS band. 

Fig.~\ref{fig:returnplots_teams} shows TEAMS gives a good estimate of the return values when all runs are pooled together, but that most individual TEAMS runs underestimate the true return values while a few overestimate them to allow for a good pooled estimate.
As in \citeA{Lucente2022coupling}, we can attribute this behavior to \emph{apparent bias}, which is best explained by analogy: an experiment consisting of 100 flips of a coin with $p=\mathbb{P}(\text{heads})=0.001$ has a nine in ten chance of landing no heads, yielding a probability estimate $\hat{p}=0$. But one experiment out of ten will yield $\hat{p}=0.01$, a gross over-estimate, and only by pooling these two scenarios together can we see the estimator's lack of bias. 
Unlike the coin-flipping experiment, TEAMS is designed to preferentially sample extreme values, but a given AMS run for L96 may still get stuck in a local optimum yielding underestimated return values, especially if the stochastic forcing is too weak to jolt a trajectory out of it. Thus, pooling over multiple runs is especially crucial in the deterministic limit. 

\section{Optimizing advance split time}
\label{sec:choosing_delta}
In this section, we explain how we determined optimal values of the advance split time $\delta$ using a simple exhaustive search. We then investigate the behavior of $\delta$ as a function of stochastic forcing strength as a guide for choosing $\delta$ prior to running TEAMS on a more expensive model for which exhaustive search would not be feasible.

\subsection{Exhaustive search}

\begin{figure}
    \centering
    \includegraphics[width=0.99\linewidth,trim={0 0cm 15cm 0cm},clip]{"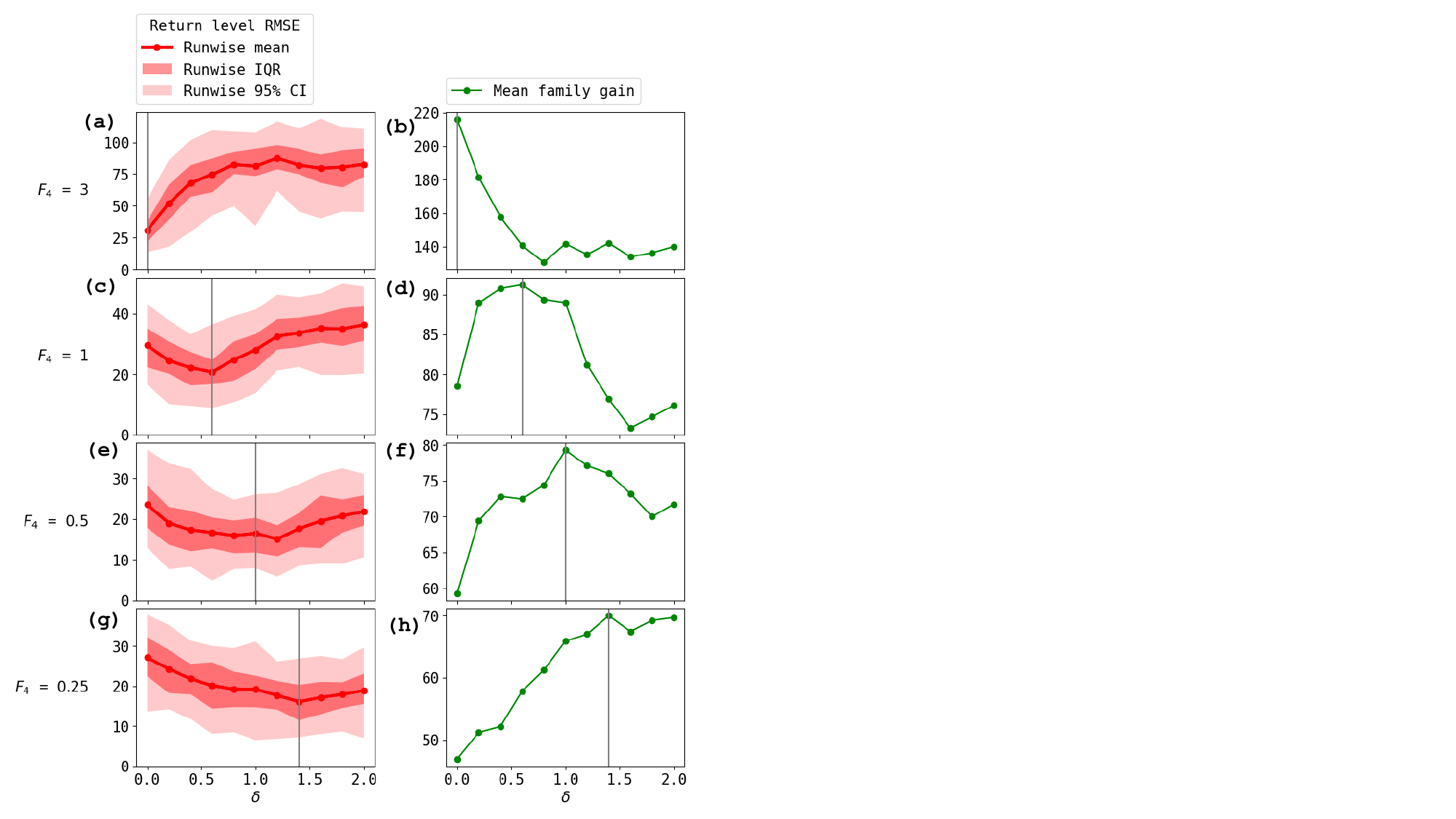"}
    \caption{Performance of TEAMS as a function of advance split time $\delta$ and as measured by (a,c,e,g) return level RMSE  and (b,d,f,h) mean family gain for $F_4=$ (a,b) 3, (c,d) 1, (e,f) 0.5, and (g,h) 0.25. Return level RMSE is computed separately for each run. Thick red lines show the average over runs, and red envelopes show the quantile ranges 25\%-75\% (or interquartile range, IQR) and 2.5\%-97.5\% across the 56 runs. Mean family gain is maximum gain in score within a single family averaged over the 56 runs.  Vertical gray lines show the optimal values of $\delta$ used in Figs.~\ref{fig:storylines_teams} and~\ref{fig:returnplots_teams}.}
    \label{fig:error_metrics}
\end{figure}

We selected the ``optimal'' $\delta$ values based on two simple performance metrics, which are plotted in Fig.~\ref{fig:error_metrics}.
\begin{enumerate}
	\item Return level RMSE: the root-mean-square difference of return level between a TEAMS estimate (from a single run) and the DNS-determined ground truth, where the mean is taken over uniform bins in $\log\tau$ space. This metric is proportional to the $L^2$-norm between a red line and the black line in the left columns of Figs.~\ref{fig:returnplots_ams} and~\ref{fig:returnplots_teams}. In cases where the red line stops before the black line, it is extrapolated to longer return periods with a constant given by its maximum to penalize the algorithm getting stuck at a false upper bound. We calculate statistics of the return level RMSE across runs, including the mean and quantiles, which are displayed in Fig.~\ref{fig:error_metrics}(a,c,e,g). Note that these correspond to \emph{percentile bootstrap} confidence intervals \cite{Wasserman2004all}, as opposed to the \emph{basic bootstrap} confidence intervals shown in Figs.~\ref{fig:returnplots_ams} and~\ref{fig:returnplots_teams}. Here we use the percentile bootstrap as a means of sensitivity analysis, to show the range of results that might occur due to sampling variability. The basic bootstrap, by contrast, is intended to bracket the ground truth of some parameter value. The return level RMSE can also be calculated for the pooled estimate, and it shows similar but noisier trends.
	\item Mean family gain: the maximum improvement (difference in scores) from ancestor to descendant over all $N$ ancestors, averaged over the 56 runs. This does not measure statistical accuracy, but only the consistent ability to generate extreme events out of moderate events. Fig.~\ref{fig:error_metrics} (b,d,f,h) shows mean family gain. Other metrics of gain, such as the maximum descendant score minus the maximum ancestral score (not necessarily from the same family tree) yield very similar trends with $\delta$, albeit different absolute values. 
\end{enumerate}
A good choice of $\delta$ should have a small return level RMSE and a large mean family gain. 
Based on both performance metrics, we selected optimal $\delta=0,0.6,1,1.4$ for $F_4=3,1,0.5,0.25$, respectively. These optimal values are marked with vertical gray lines in Fig.~\ref{fig:error_metrics}, and they are used in Figs.~\ref{fig:storylines_teams} and~\ref{fig:returnplots_teams}.  
For $F_4=0.5$, the two metrics gave slightly difference optimal values ( $\delta=1.2$ for return level RMSE or $\delta=1$ for mean family gain); we chose $\delta=1$ because it gave the better pooled estimate.
We emphasize that the optimal values are only discernible by averaging over many independent runs. 
For completeness, we display all 44 return level vs. return period plots (4 values of $F_4$ $\times$ 11 values of $\delta$) in the supplement. 
In general, shifting the optimal $\delta$ by $\pm0.2$ doesn't change the results qualitatively, but larger shifts can degrade performance.   
The absolute values of errors should not be compared between stochastic forcing levels, since each has its own statistical steady state. Rather, the important takeaway is the increase in optimal $\delta$ as the stochastic forcing weakens.
Indeed, in the singular limit of zero stochastic forcing the advance split time must necessarily go to infinity to have any effect at all, and initial condition perturbations would be needed to split trajectories.

To summarize, we have found that some choices of $\delta$ can make TEAMS effective where AMS is not effective, and that the optimal $\delta$ increases as stochastic forcing magnitude decreases. In the next section we relate this behavior to the predictability time, which points toward a cheap method to estimate an optimal---or at least, reasonably performant---$\delta$, without having to repeatedly run TEAMS.

\subsection{Relation between optimal advance time and error saturation timescales}
\label{sec:prt}

Heuristically, we expect the optimal advance time $\delta$ to reflect the divergence timescale of perturbed trajectories that are introduced in splitting. Can this be related to classical predictability timescales? Lyapunov analysis describes perturbation growth by way of Lyapunov exponents and singular vectors 
\cite{Cencini2013lyapunov,Norwood2013lyapunov,
Pazo2010spatiotemporal,Maiocchi2024heterogeneity}, but it 
applies to the regime of \emph{infinitesimal} perturbations. 
The kind of perturbations we strive for in rare event sampling are finite and nonlinear, turning peaks into substantially larger peaks as in Figs.~\ref{fig:storylines_ams},~\ref{fig:storylines_teams}. 
``Finite size Lyapunov exponents'' (FSLEs)  \cite{Boffetta1998extension,Cencini2013finite} are closer to what we need, generalizing the Lyapunov exponent to be dependent on an initial error amplitude. Typically, error grows in two stages: first exponentially, during which the FSLE equals the leading Lyapunov exponent, and then diffusively (scaling as a power law with time), during which the FSLE declines. The divergence timescale is bounded below by this change point, which approaches zero as stochastic forcing becomes dominant: indeed, the variance of pure Brownian motion grows linearly in $t$ immediately.

On the other hand, the optimal $\delta$  is bounded above by the \emph{error saturation timescale}, when perturbed ensemble members become independent and inhabit totally different regions of the attractor. By then, the root-mean-square error (RMSE) of the ensemble will equal the root-mean-square distance (RMSD) between two randomly chosen points on the attractor. In climate models, the saturation timescale is a convenient and effective measure of predictability \cite{Sheshadri2021midlatitude}. Clearly, $\delta$ must be chosen shorter than the time to saturation, since a child trajectory ought to take advantage of pre-existing maxima produced by its parent. To investigate this relationship, the following experiments measure time in terms of fraction of saturation. 

For each $F_4$ considered, we ran a moderate-length control simulation $x(t)$ for $0\leq t\leq 1050$ (discarding the first 50 as spinup), and measured the RMSD for this simulation. At initialization times $50,70,90,...,990$ (48 in total) we branched a 16-member ensemble with identical initial conditions $x(t)$ but independent stochastic forcing realizations (a convenient feature of stochastic forcing is that errors grow even from perfect initial conditions, removing dependence on initial perturbation amplitude). We integrated each member for 15 time units, calculated RMSE as a function of time (separately for each ensemble), and inverted to find the times $t_\epsilon$ at which the fraction of saturation $\epsilon = $RMSE/RMSD reached a given value. In other words, $\mathrm{RMSE}(t_\epsilon)=\epsilon \times \mathrm{RMSD}$. Finally, we take the average across initializations to get a single value $\overline{t_\epsilon}$ for each of several $\epsilon$ values. The total cost of this experiment is $1.2\times10^4$ time units, roughly equal to 1.5 runs of AMS and much cheaper than the 56-run pooled estimate. Moreover, halving the number of initializations used yields qualitatively similar results.

\begin{figure}
    \centering
    \includegraphics[width=0.99\linewidth,trim={0 2cm 0cm 0cm},clip]{"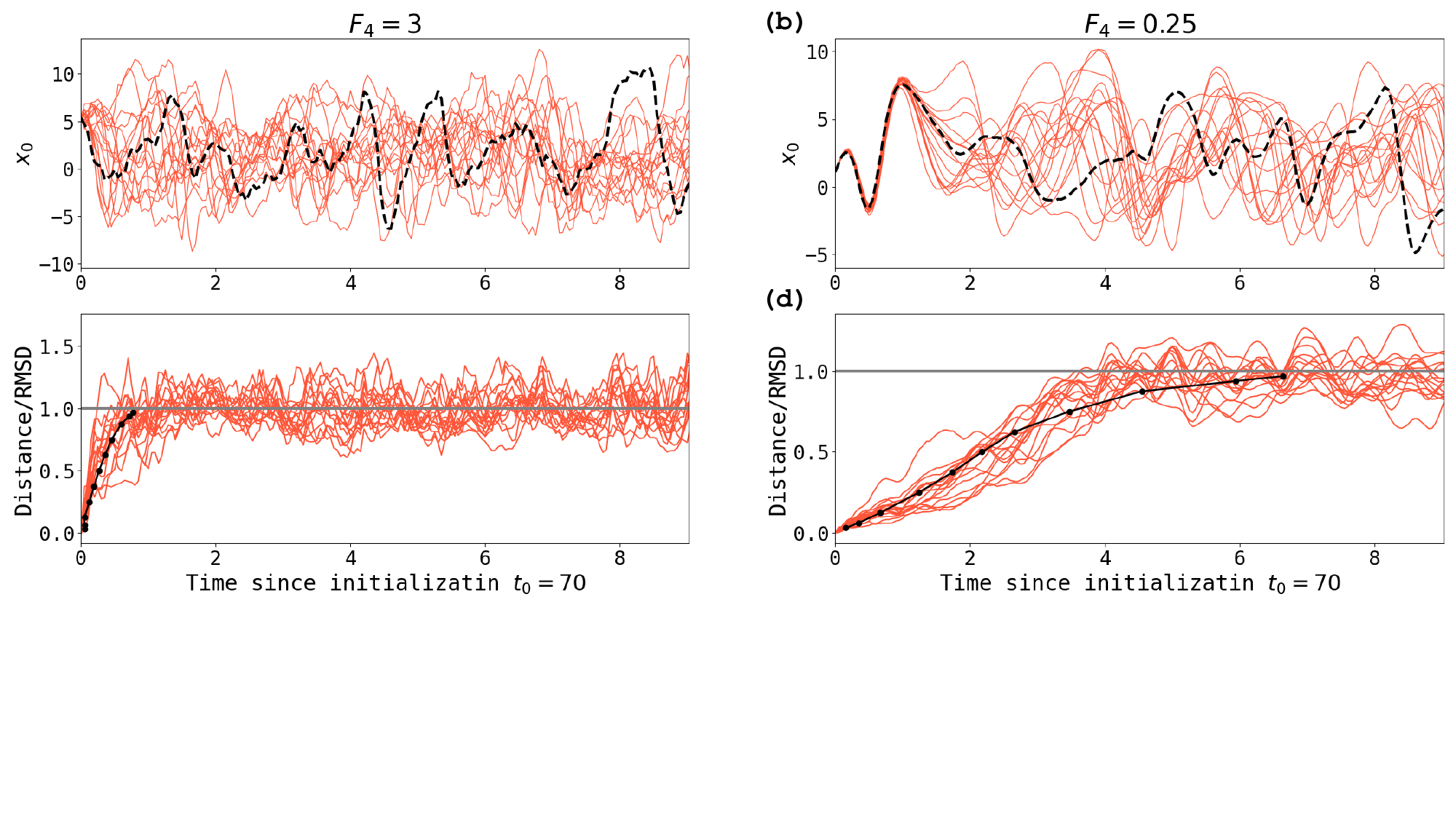"}
    \caption{Growth of perturbations in the experiments described in subsection~\ref{sec:prt} for one representative initialization time $t_0=70$ and two values of the stochastic forcing: (a,c) $F_4=3$ and (b,d) $F_4=0.25$ . (a,b) show $x_0(t)$ for the control simulation (black) and 16 simulations with the same initial condition but different white-noise forcing realizations (red). (c,d) show Euclidean distance between each ensemble member to the control as a fraction of RMSD versus time (red), and the fraction of saturation RMSE/RMSD versus the time until each $\epsilon$ value is reached averaged across all initializations and ensemble members (black), i.e., $\epsilon$ vs. $\overline{t_\epsilon}$. Dots indicate $\epsilon=1/32,1/16,1/8,1/4,3/8,1/2$, and these same values reflected about $1/2$.}
    \label{fig:pert_spaghetti}
\end{figure}

Fig.~\ref{fig:pert_spaghetti} shows timeseries of $x_0(t)$ (both control and perturbed) and error growth for two such ensembles from the high and low stochastic forcing cases.
The time axis is truncated to 10 days past initialization. 
The early linear growth of $\epsilon$ vs. $\overline{t_\epsilon}$ indicates a steady decline in relative growth rate, meaning that the perturbations begin to enter the diffusive (sub-exponential) growth regime quite early. This is thanks to stochastic forcing, which is visible in the top row as the emergence of red members from the shadow of the control trajectory. As expected, the error growth is much faster for the higher value of stochastic forcing.
%

\begin{figure}
    \centering
    \includegraphics[width=0.6\linewidth,trim={0 0cm 6cm 0cm},clip]{"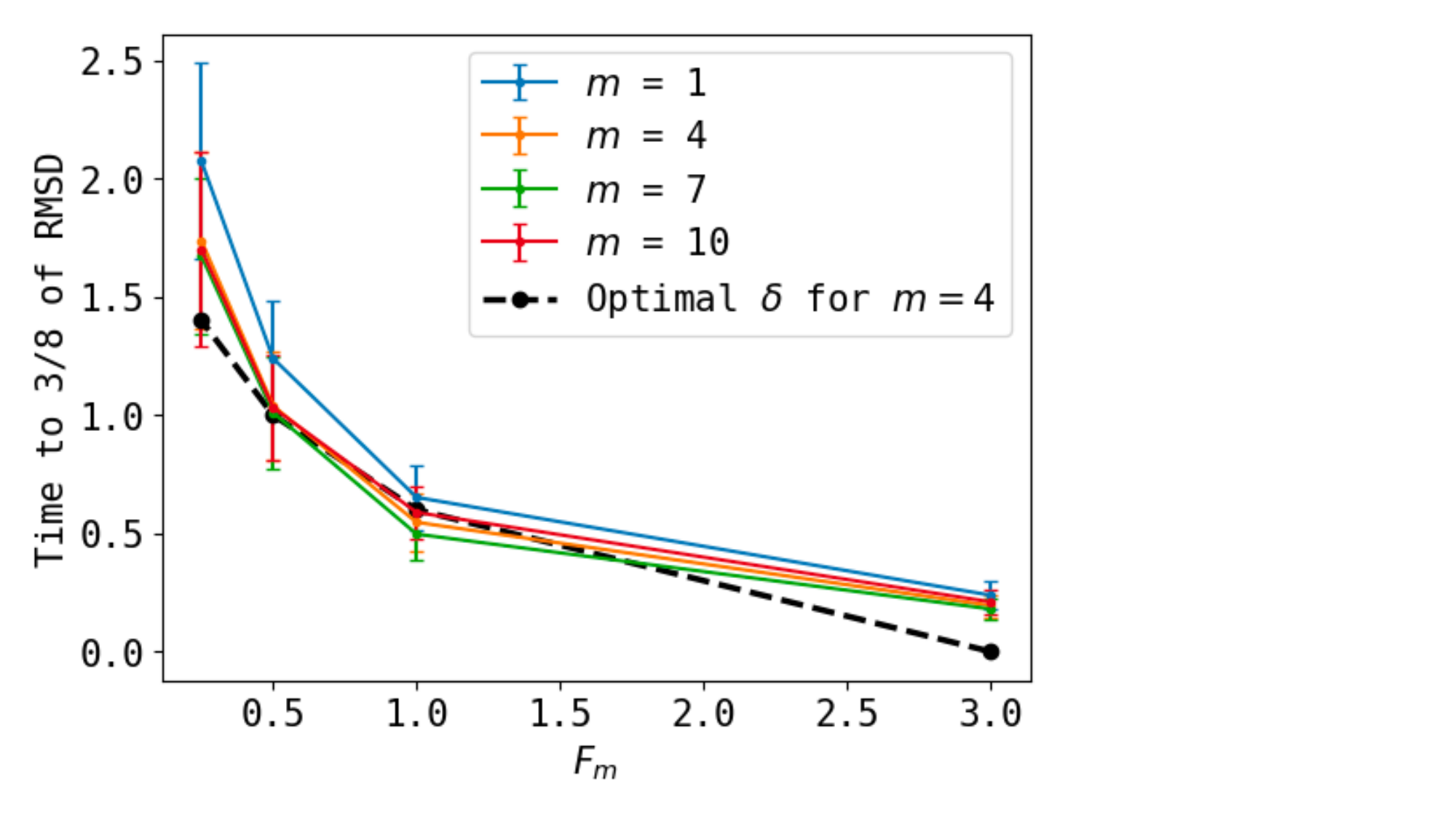"}
    \caption{Time $\overline{t_{3/8}}$ until the perturbations described in subsection~\ref{sec:prt} reach a fixed fraction (3/8) of RMSD as a function of stochastic forcing strength $F_m$ for different wavenumbers $m$. Error bars are $\pm1$ standard deviation of the distribution over different initial conditions. Optimized values of $\delta$ (determined from the performance metrics in Fig.~\ref{fig:error_metrics}) are shown in the black dashed line for $m=4$.}
    \label{fig:sat_times}
\end{figure}

If the optimal $\delta$ could be predicted from the error growth rates alone, the TEAMS algorithm could be calibrated simply and cheaply before being deployed. 
Fig.~\ref{fig:sat_times}  shows the time $\overline{t_{3/8}}$ when RMSE reaches a fixed fraction of RMSD ($3/8$) as compared to the optimal $\delta$ values determined from Fig.~\ref{fig:error_metrics}, as a function of the strength of stochastic forcing. We include results from forcing at wavenumbers $m=1,4,7,10$.  
There is an encouraging similarity between the dependence of optimal $\delta$ and $\overline{t_{3/8}}$ on stochastic forcing strength, suggesting that the fractional saturation time might be useful to provide an estimate for $\delta$.

Another interesting and less obvious feature is the dependence on wavenumber of error growth (albeit a weak dependence): medium-length wave forcing ($m=4$ and $m=7$) drives error to saturation faster than very short ($m=10$) or long ($m=1$) wave forcing, which informed our choice of $m=4$ throughout the TEAMS experiments. However, the variability due to initial conditions (indicated by $\pm1\sigma$ error bars) tend to exceed systematic differences between wavenumbers. This variability reflects a distribution of divergence timescales across the attractor, which was also found be be quite heterogeneous in \citeA{Maiocchi2024heterogeneity} (there measured by Lyapunov exponents). It also suggests that the best strategy may be to not fix a single $\delta$, but to allow the algorithm to adaptively set a $\delta$, or sample from a range, to account for differing divergence timescales between different initial conditions,  and this could be investigated in future work.

\section{Conclusions and Outlook}
\label{sec:conclusion}

A vexing challenge in climate science is reliably quantifying the probability of extreme weather events, which are fundamentally difficult to characterize because of data scarcity. Among various competing strategies, rare event algorithms hold several key advantages, chiefly (i) access to dynamical samples of the events, rather than just return period curves which extreme value theory might provide, and (ii) more statistical rigor than storyline-based approaches like ``ensemble boosting'' \cite{Gessner2021very}, thanks to careful re-weighting of cloned trajectories. Inspired by recent successes of rare event algorithms on long-lasting heat waves \cite{Ragone2018computation} and idealized models of regime transitions \cite{Lucente2022coupling,JacquesDumas2023data}, we have investigated the ability of a particular algorithm, adaptive multilevel splitting (AMS) to sample extreme events of a different character: intermittent, short-lived bursts of energy in the Lorenz-96 model which have some similar characteristics as extreme daily rain or wind associated with midlatitude cyclones.

Even in this simple model, we have elucidated some key obstacles that hinder rare event splitting algorithms on sudden, short-lived events, and furthermore taken some steps to overcome them. AMS sets up a sequence of thresholds for an observable of interest and estimates conditional exceedance probabilities in stages by cloning and perturbing ``successful'' ensemble members when they cross a threshold, to generate new ``successful'' samples. This simple prescription suffers a fatal problem when the events are short-lived compared to the divergence timescale (how long it takes a perturbation to grow appreciably): a perturbed ensemble member essentially replicates its parent's success, and doesn't develop its own history until after the event is over. Neither the magnitude nor the diversity of rare event samples is enhanced. To fix this problem, we have drawn inspiration from ensemble boosting to apply a perturbation \emph{in advance} of the rare event by some lead time $\delta$. But we have also retained rigorous statistics for these ``storylines'' by exploiting a more general rare event algorithm, subset simulation (SS), of which AMS is only a special case. We name the resulting algorithm ``trying-early AMS'' (TEAMS) and demonstrate its success in sampling the tails of the rare event distribution more efficiently than direct numerical simulation can do, despite an extra computational cost due to rejected samples.

Our study is a proof of concept that suggests a path forward, but with some open questions and directions for improvement, which we summarize here: 
\begin{itemize}
	\item The most crucial algorithmic parameter is the advance split time, $\delta$, which is equivalent to a proposal distribution width. Our grid search for optimal $\delta$, though not a scalable solution, demonstrates a relationship with the time over which perturbations grow to a fraction of saturation. An important goal for future work is to assess this result for other underlying models such as general circulation models or for other error growth metrics. Given the localized nature of our observable ($x_0^2$ is the energy at a single longitude site), it is interesting that a \emph{global} Euclidean metric correlates with the optimal $\delta$. Weighting the metric more heavily for grid points near $k=0$ might further improve this relationship.  
	\item The weak stochastic forcing limit $F_m\to0$ is important to confront for climate models, which may be more practical to perturb just at the splitting time rather than continuously at every time step, especially if the climate model is not already equipped with a stochastic subgrid parameterization. In the TEAMS framework, this would translate to perturbing a simulation at a lead time $\delta$ ahead of the event, but not at all following times. Perturbing at just one time makes a given perturbation magnitude less powerful---but also opens up interesting possibilities such as the use of deterministic optimization strategies to more efficiently discover the most extreme event possible from a given initial condition. For example, some directions of perturbation (singular vectors) grow much faster than others, a fact which has informed ensemble design in operational weather forecasting \cite{Palmer2013singular}, and could be used to further improve the algorithm. Methods such as conditional nonlinear optimal perturbation \cite<>[and references therein]{Wang2020useful}, generalized stability theory \cite{Farrell1996generalized}, and large deviation theory \cite{Dematteis2017rogue,Dematteis2019extreme,Schorlepp2023scalable} may prove useful for this task. 
	\item Related to the previous point, it is desirable to have greater efficiency with samples in order to deploy rare event algorithms at scale. For example, we should not simply discard rejected samples, but rather learn from their ``mistakes'' to design better perturbations. Frameworks like Bayesian optimization and adaptive importance sampling based on model reduction have been developed for this task, and have been used in safety assessment for reliability engineering \cite<e.g.,>{Cousins2014quantification,Huang2016assessing,Mohamad2018sequential,Sapsis2020output,Uribe2021cross,Zhang2022koopman}. 
\end{itemize}

Rare event algorithms represent a new way to allocate computational resources to where they matter most. To realize their considerable potential for efficiency gains, we have taken one of the  necessary steps to make them flexible enough to target intermittent, localized, transient events that characterize phenomena such as heavy precipitation in complex global climate models. The Lorenz-96 model is an invaluable prototype as a cheap system that poses similar algorithmic challenges. Forthcoming papers will use the insight gained here as a stepping stone to more complex and realistic models.

\section*{Data availability statement}
The software to simulate and sample extreme events in Lorenz-96 using TEAMS is available in a public Zenodo repository at https://zenodo.org/doi/10.5281/zenodo.10608187. Interested readers are encouraged to try out the algorithm on other systems of interest, and should not hesitate to contact J. F. for assistance.

\acknowledgments
This research is part of the MIT Climate Grand Challenge on Weather and Climate Extremes. It received support by the generosity of Eric and Wendy Schmidt by recommendation of Schmidt Sciences as part of its Virtual Earth System Research Institute (VESRI).
Computations for this research were carried out on the MIT Engaging cluster. 

%
%

\bibliography{references_cap}

%
%
%
%
%


\pagebreak

\section*{Supporting Information for ``Bringing statistics to storylines: rare event sampling for sudden, transient extreme events''}

\noindent\textbf{Contents}
\begin{enumerate}
\item Figures~\ref{fig:storylines_ams_ou} to~\ref{fig:delta2p0}
\end{enumerate}

\setcounter{figure}{0}
\renewcommand{\thefigure}{S\arabic{figure}}

\noindent\textbf{Introduction}
Figs.~\ref{fig:storylines_ams_ou} and~\ref{fig:ams_ou_returnplot} display results for AMS applied to the stochastic L96 model with $a=0$ and $F_4=3$, which is really just an array of correlated OU processes with no advection. Figs.~\ref{fig:delta0p0}-\ref{fig:delta2p0} display return level vs. return period plots for all combinations of stochastic forcing level $F_4\in\{3,1,0.5,0.25\}$ and the advance splitting time $\delta\in\{0,0.2,0.4,...,2\}$, only a subset of which are shown in the main text. 

\pagebreak

\begin{figure}
    \centering
    \includegraphics[width=0.8\linewidth,trim={0cm 1cm 0cm 0cm},clip]{"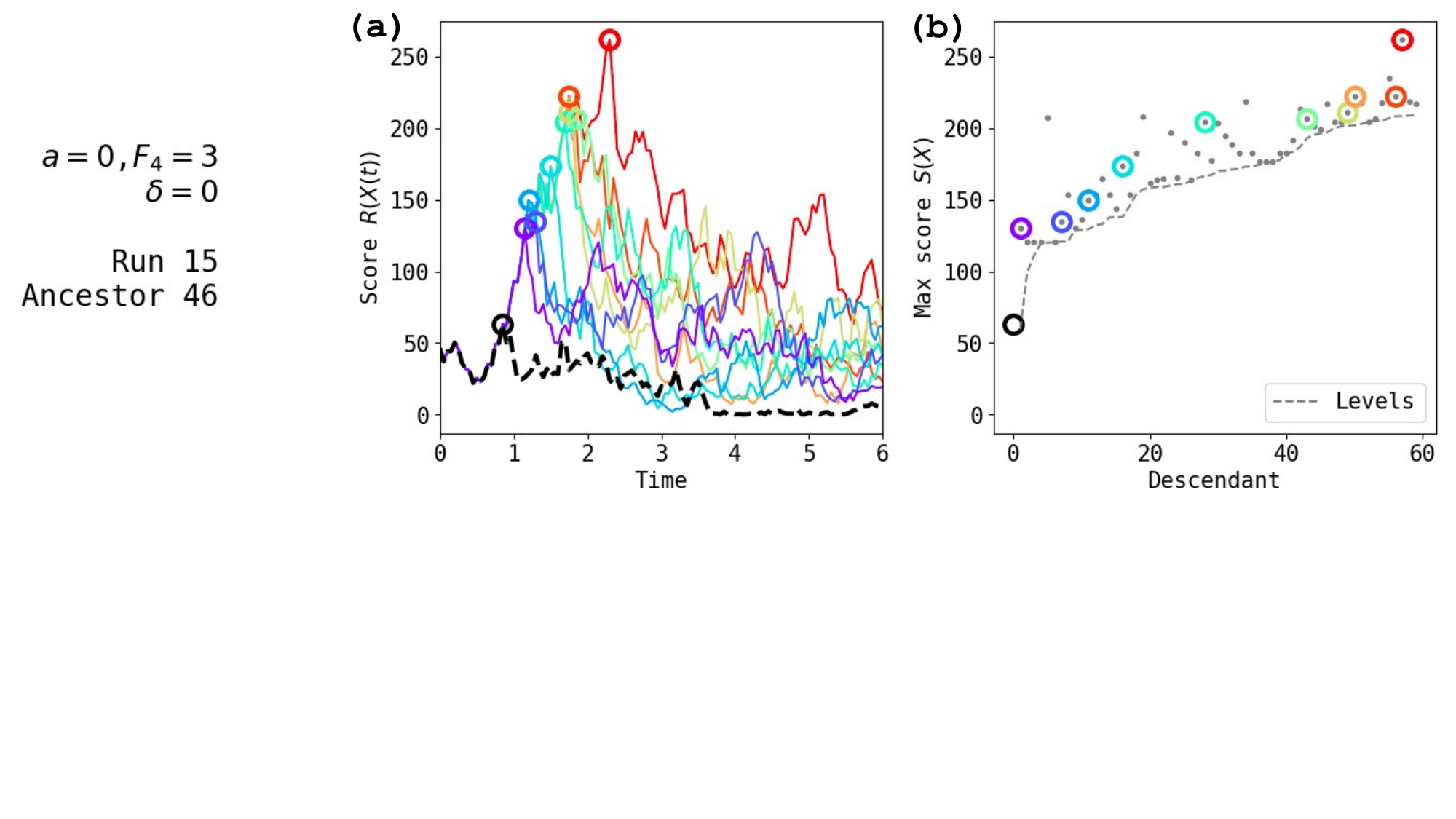"}
    \caption{Example of a single lineage generated by AMS applied to the the OU process (L95 with $a=0, F_4=3$), formatted the same as Fig. 4a,b of the main text.}
    \label{fig:storylines_ams_ou}
\end{figure}

\begin{figure}
    \centering
    \includegraphics[width=0.99\linewidth,trim={0cm 8cm 0cm 0cm},clip]{"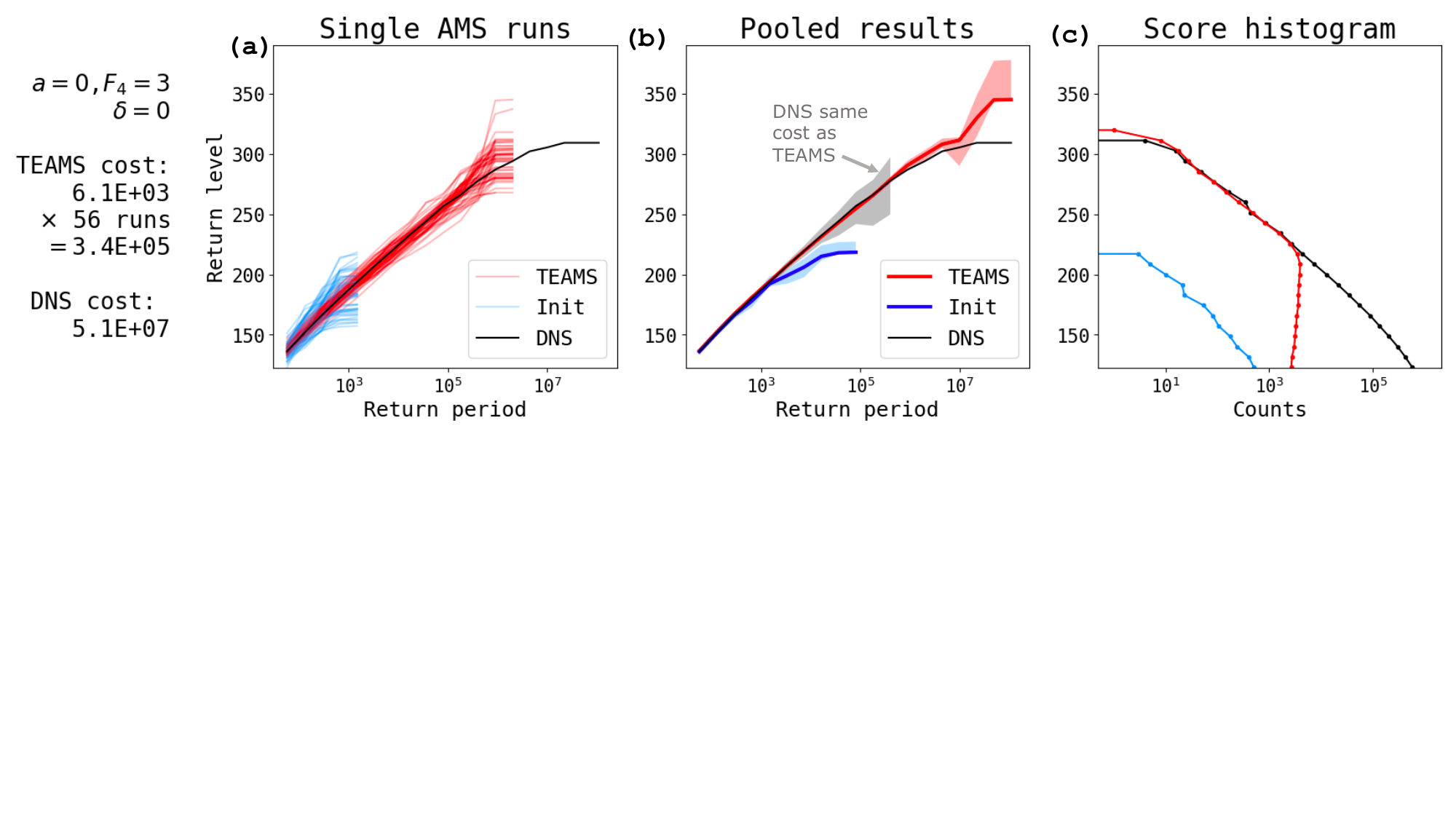"}
    \caption{
    Statistical results of AMS applied to the OU process (L96 with $a=0,F_4=3$) with $N=128$ initial ensemble members and $M=56$ runs. Format is the same as Fig. 5a,b,c of the main text.
    }
    \label{fig:ams_ou_returnplot}
\end{figure}

\begin{figure}
    \includegraphics[width=0.8\linewidth,trim={8cm 0cm 8cm 0cm},clip]{"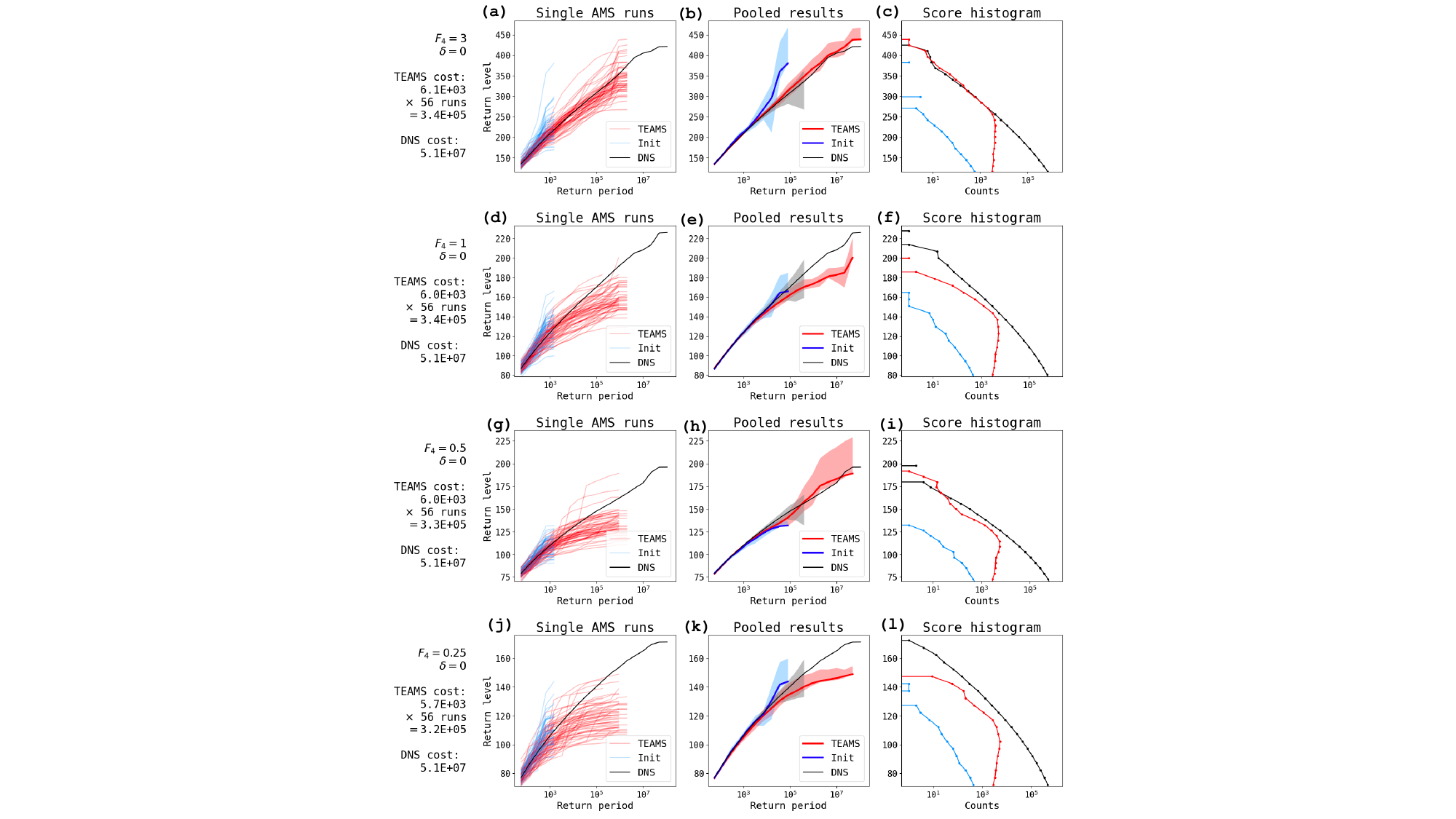"}
    \caption{TEAMS algorithm performance at all four noise levels with $\delta=0.0$}
    \label{fig:delta0p0}
\end{figure}

\begin{figure}
    \includegraphics[width=0.8\linewidth,trim={8cm 0cm 8cm 0cm},clip]{"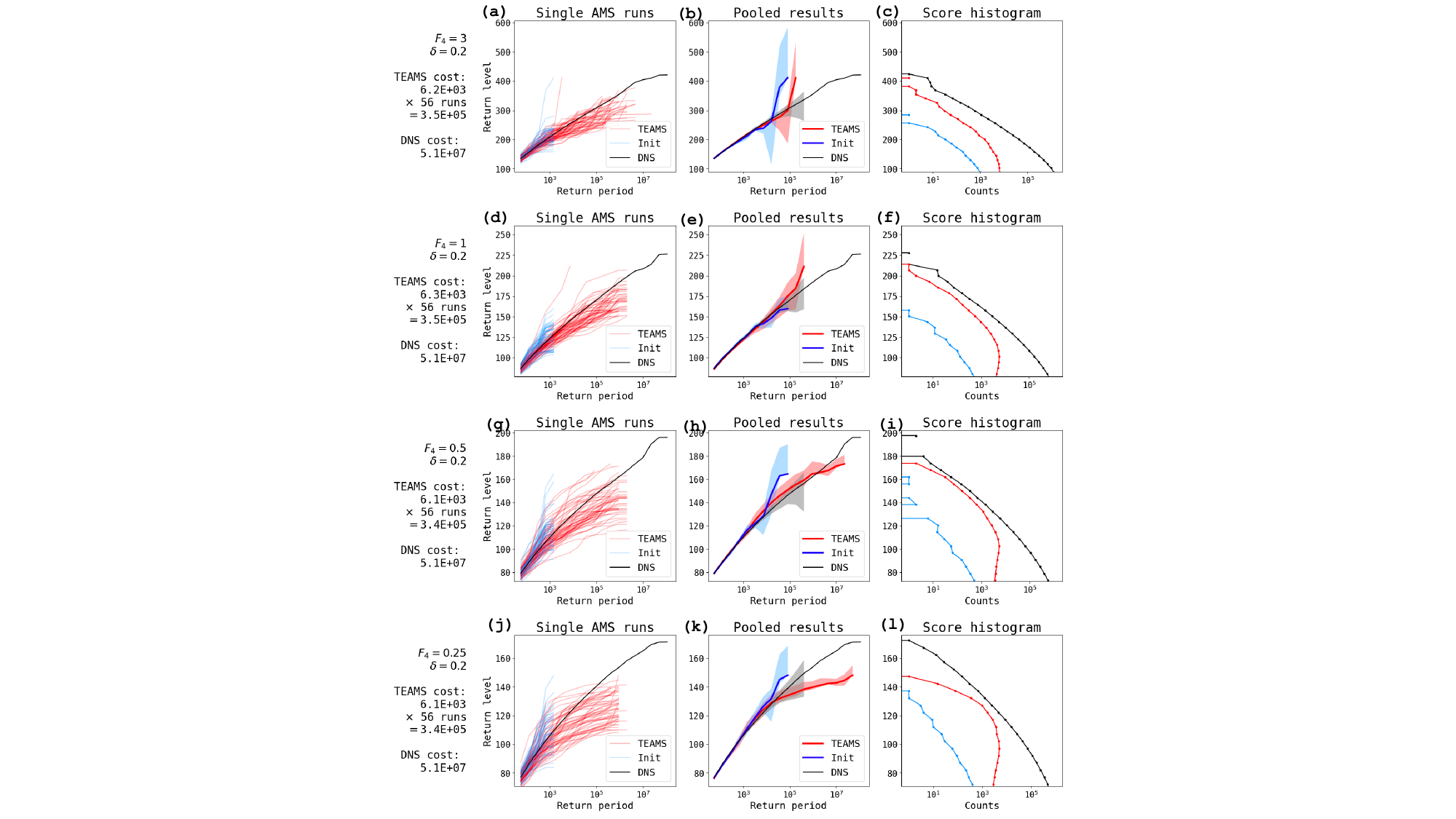"}
    \caption{TEAMS algorithm performance at all four noise levels with $\delta=0.2$}
    \label{fig:delta0p2}
\end{figure}

\begin{figure}
    \includegraphics[width=0.8\linewidth,trim={8cm 0cm 8cm 0cm},clip]{"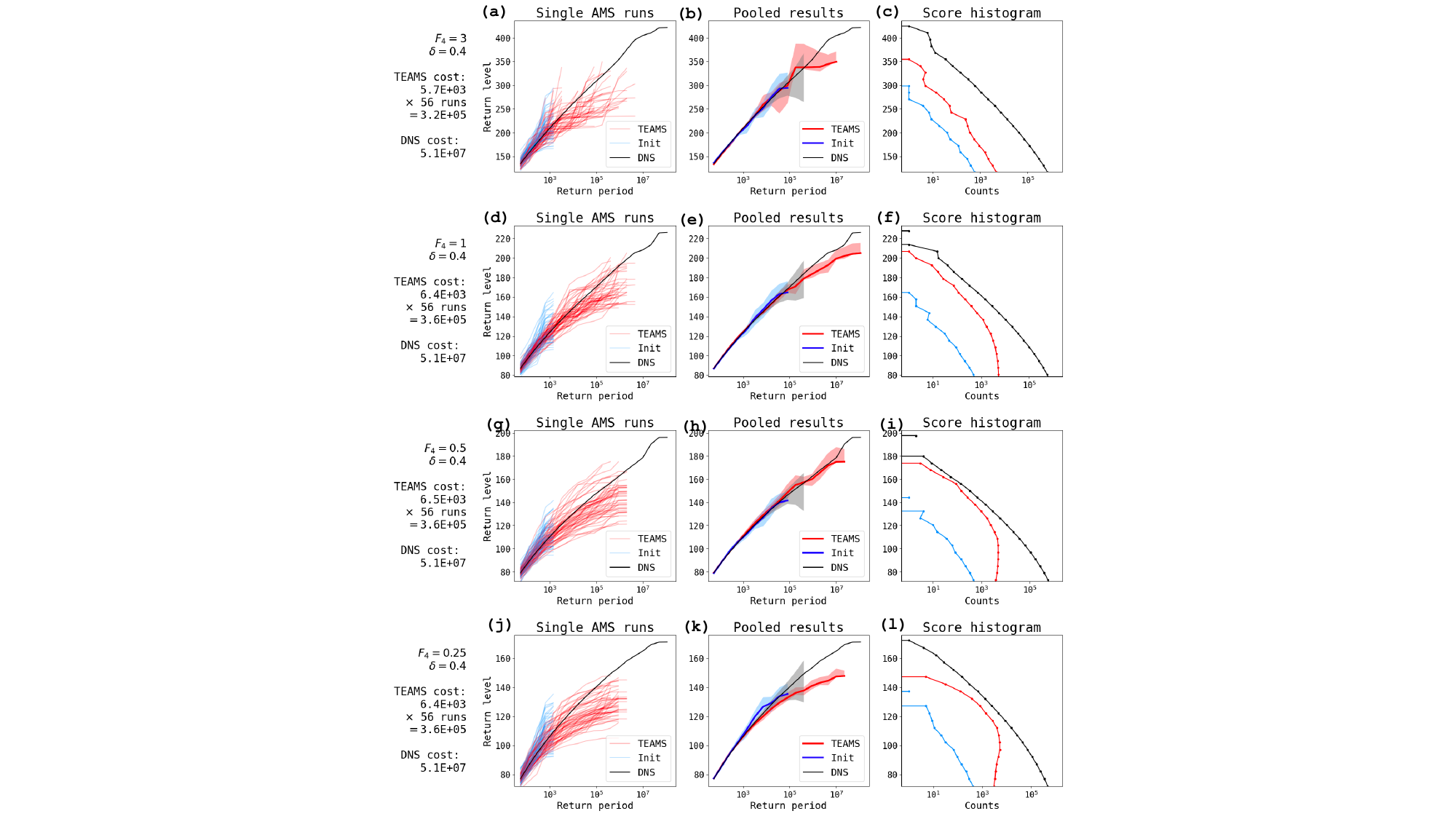"}
    \caption{TEAMS algorithm performance at all four noise levels with $\delta=0.4$}
    \label{fig:delta0p4}
\end{figure}

\begin{figure}
    \includegraphics[width=0.8\linewidth,trim={8cm 0cm 8cm 0cm},clip]{"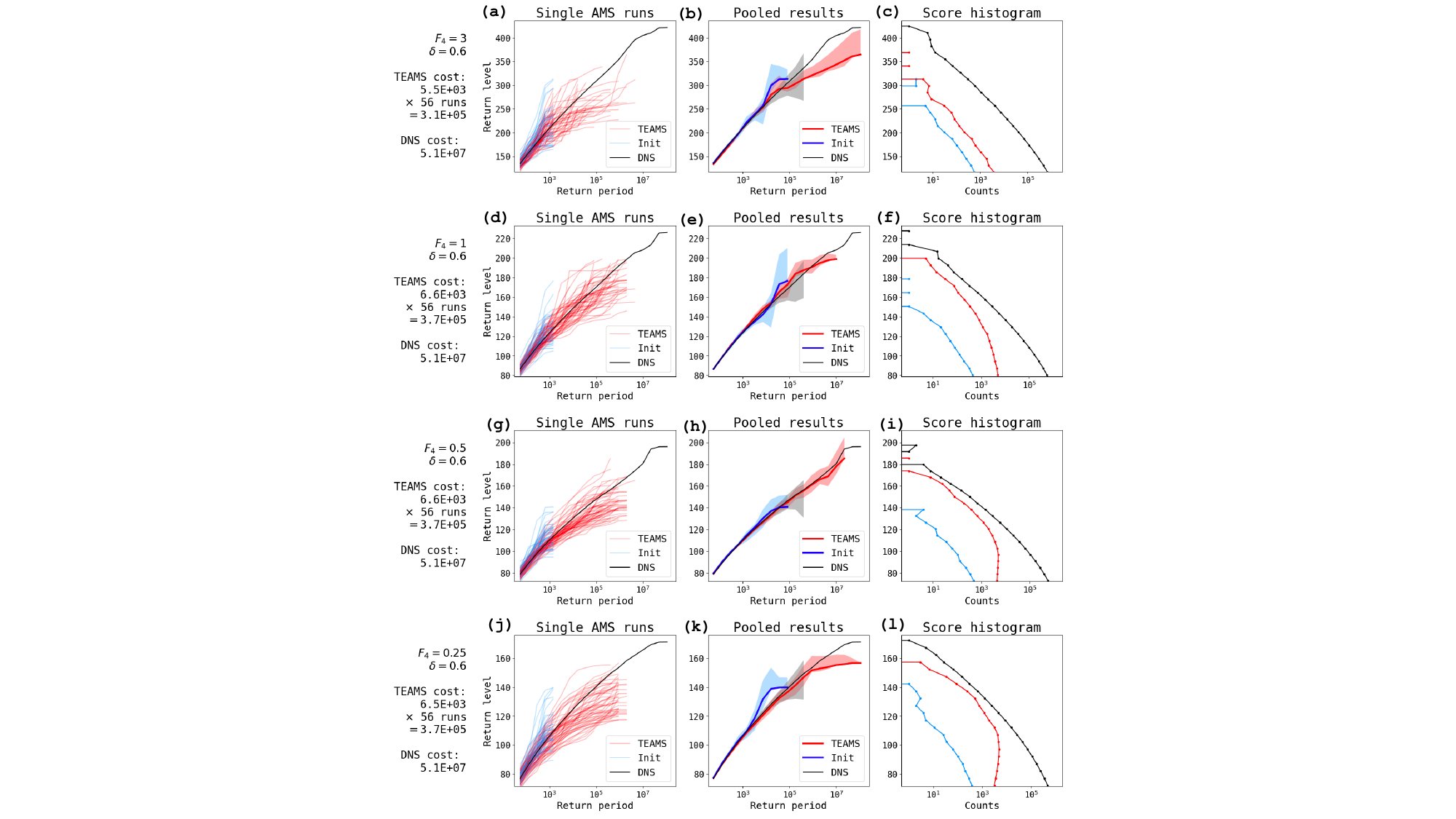"}
    \caption{TEAMS algorithm performance at all four noise levels with $\delta=0.6$}
    \label{fig:delta0p6}
\end{figure}

\begin{figure}
    \includegraphics[width=0.8\linewidth,trim={8cm 0cm 8cm 0cm},clip]{"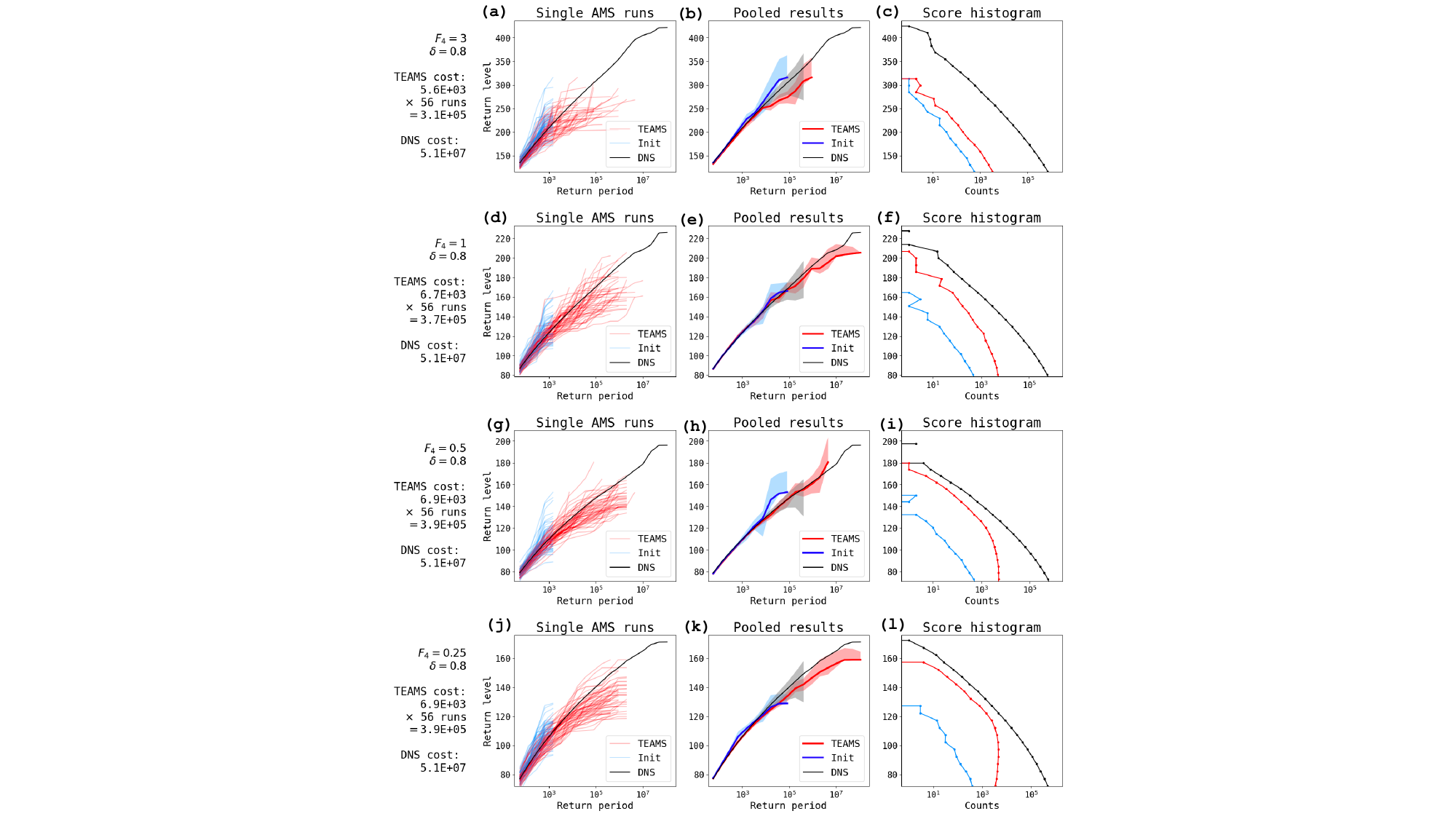"}
    \caption{TEAMS algorithm performance at all four noise levels with $\delta=0.8$}
    \label{fig:delta0p8}
\end{figure}

\begin{figure}
    \includegraphics[width=0.8\linewidth,trim={8cm 0cm 8cm 0cm},clip]{"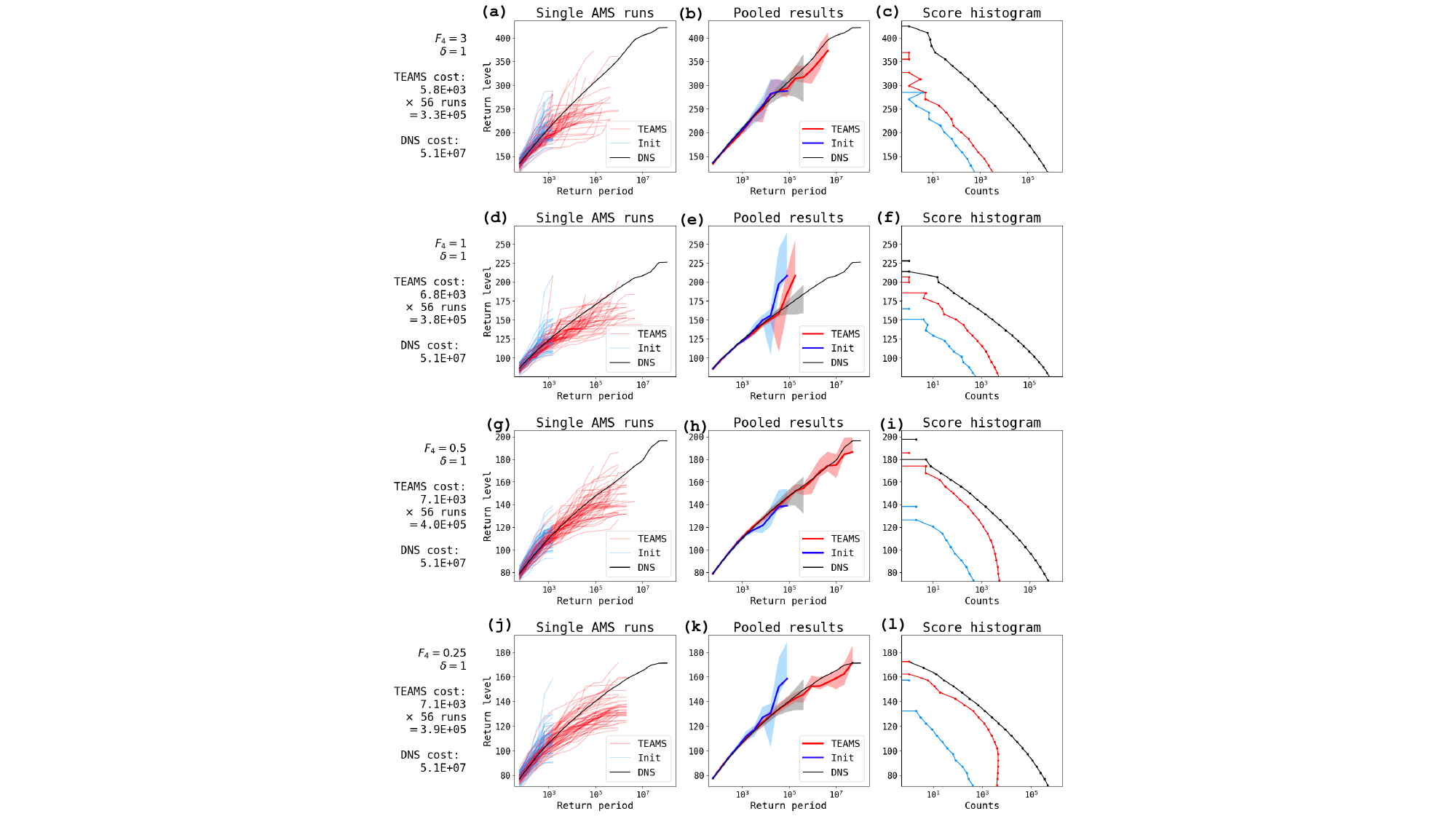"}
    \caption{TEAMS algorithm performance at all four noise levels with $\delta=1.0$}
    \label{fig:delta1p0}
\end{figure}

\begin{figure}
    \includegraphics[width=0.8\linewidth,trim={8cm 0cm 8cm 0cm},clip]{"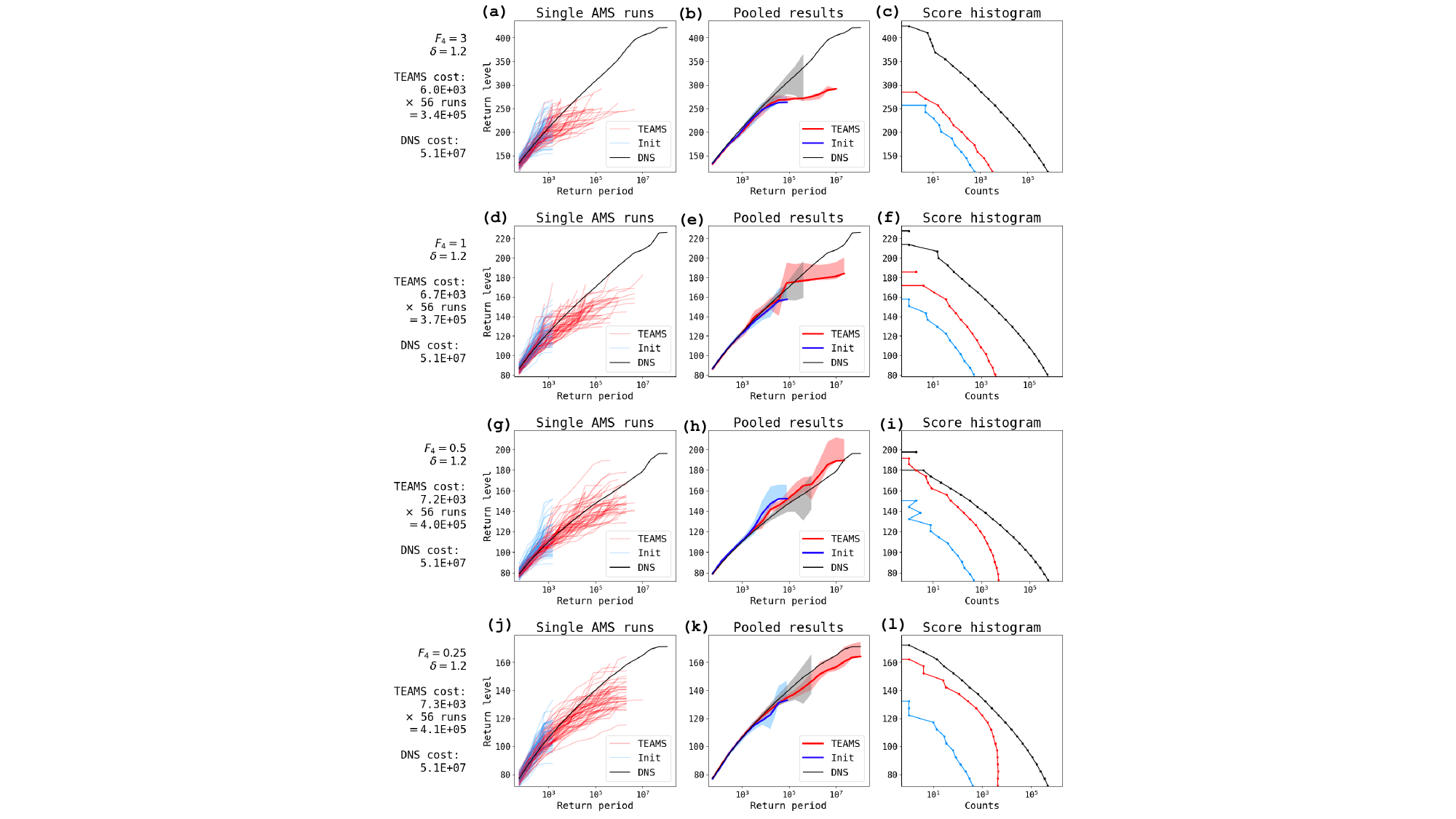"}
    \caption{TEAMS algorithm performance at all four noise levels with $\delta=1.2$}
    \label{fig:delta1p2}
\end{figure}

\begin{figure}
    \includegraphics[width=0.8\linewidth,trim={8cm 0cm 8cm 0cm},clip]{"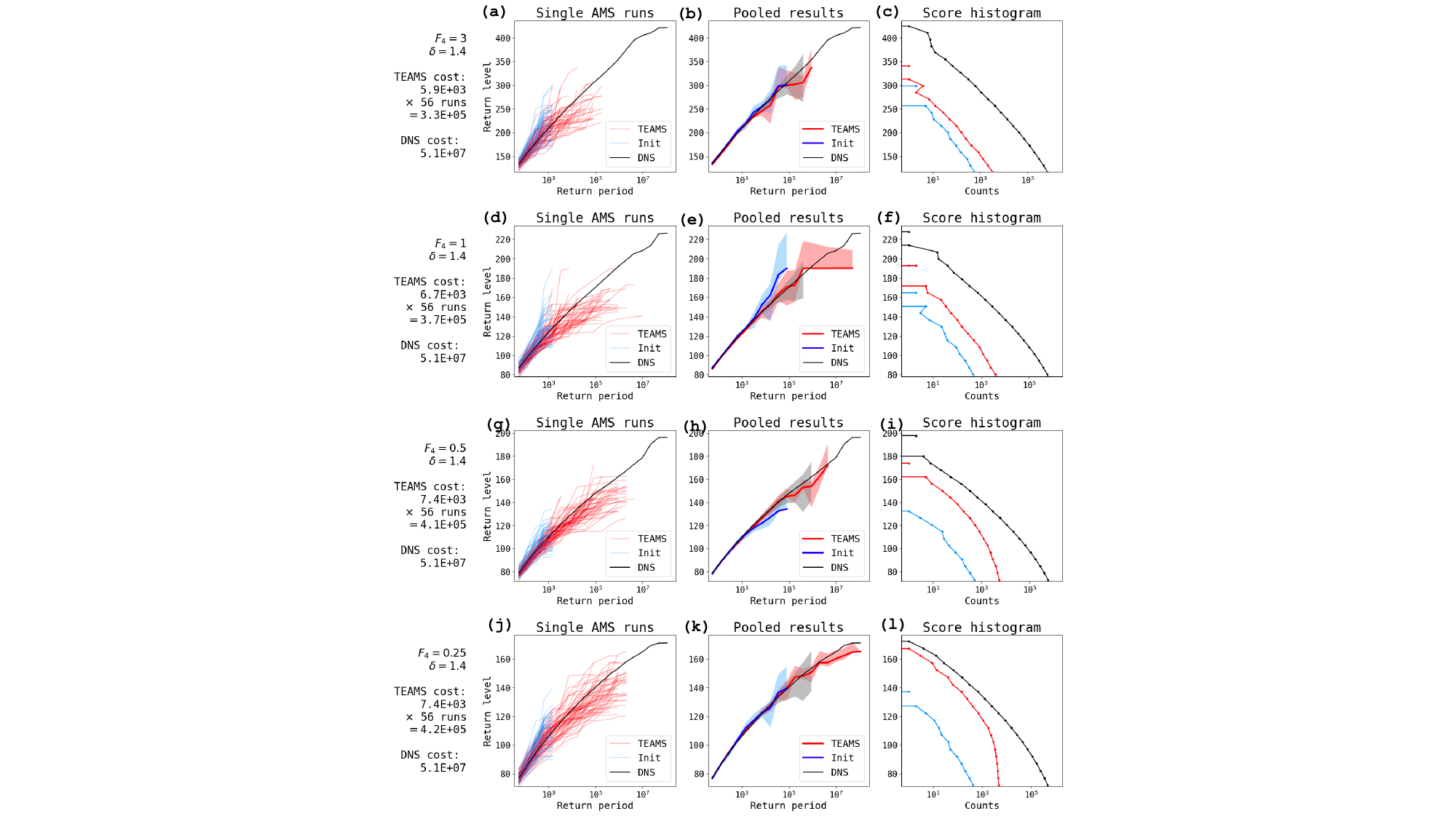"}
    \caption{TEAMS algorithm performance at all four noise levels with $\delta=1.4$}
    \label{fig:delta1p4}
\end{figure}

\begin{figure}
    \includegraphics[width=0.8\linewidth,trim={8cm 0cm 8cm 0cm},clip]{"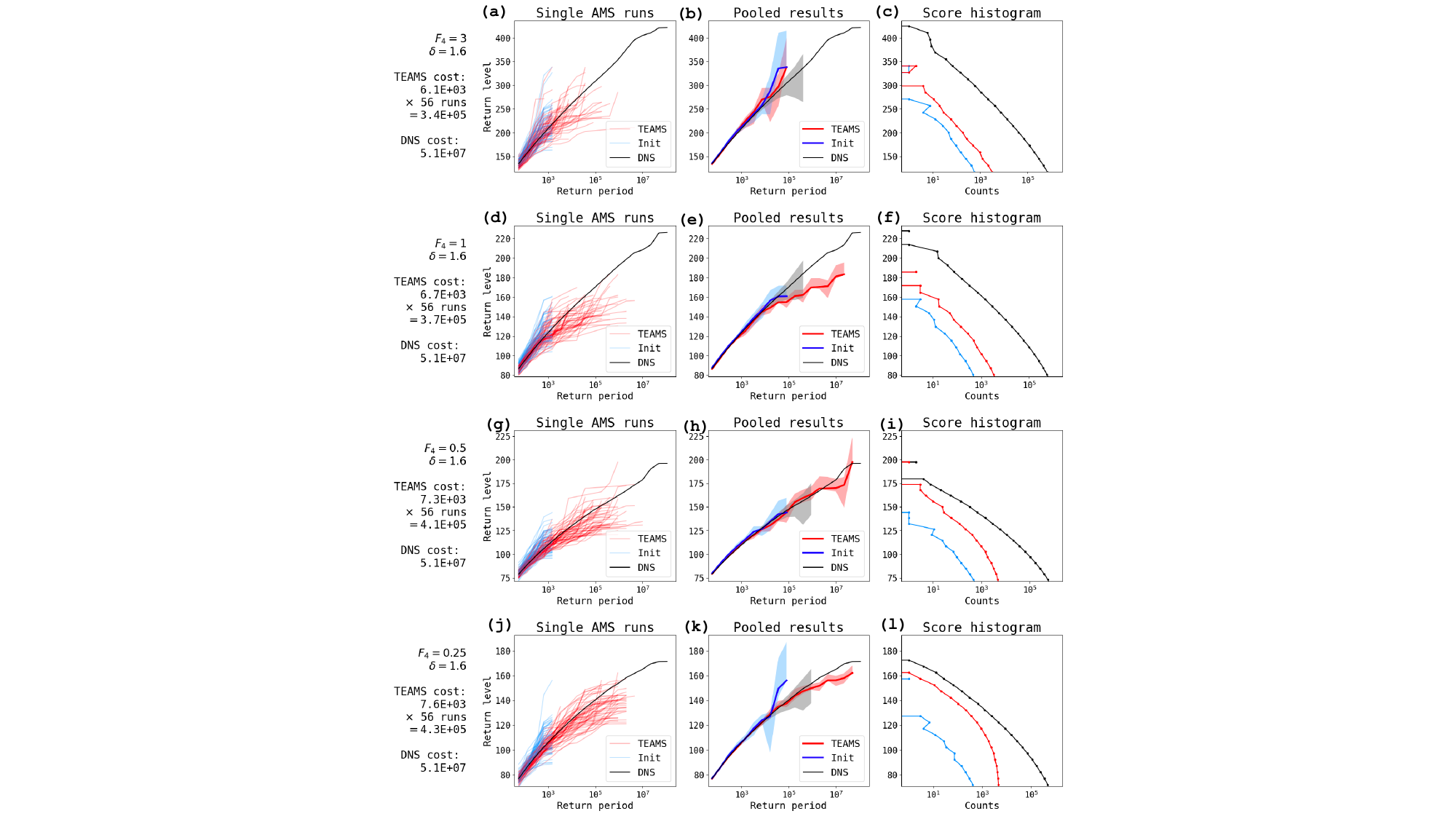"}
    \caption{TEAMS algorithm performance at all four noise levels with $\delta=1.6$}
    \label{fig:delta1p6}
\end{figure}

\begin{figure}
    \includegraphics[width=0.8\linewidth,trim={8cm 0cm 8cm 0cm},clip]{"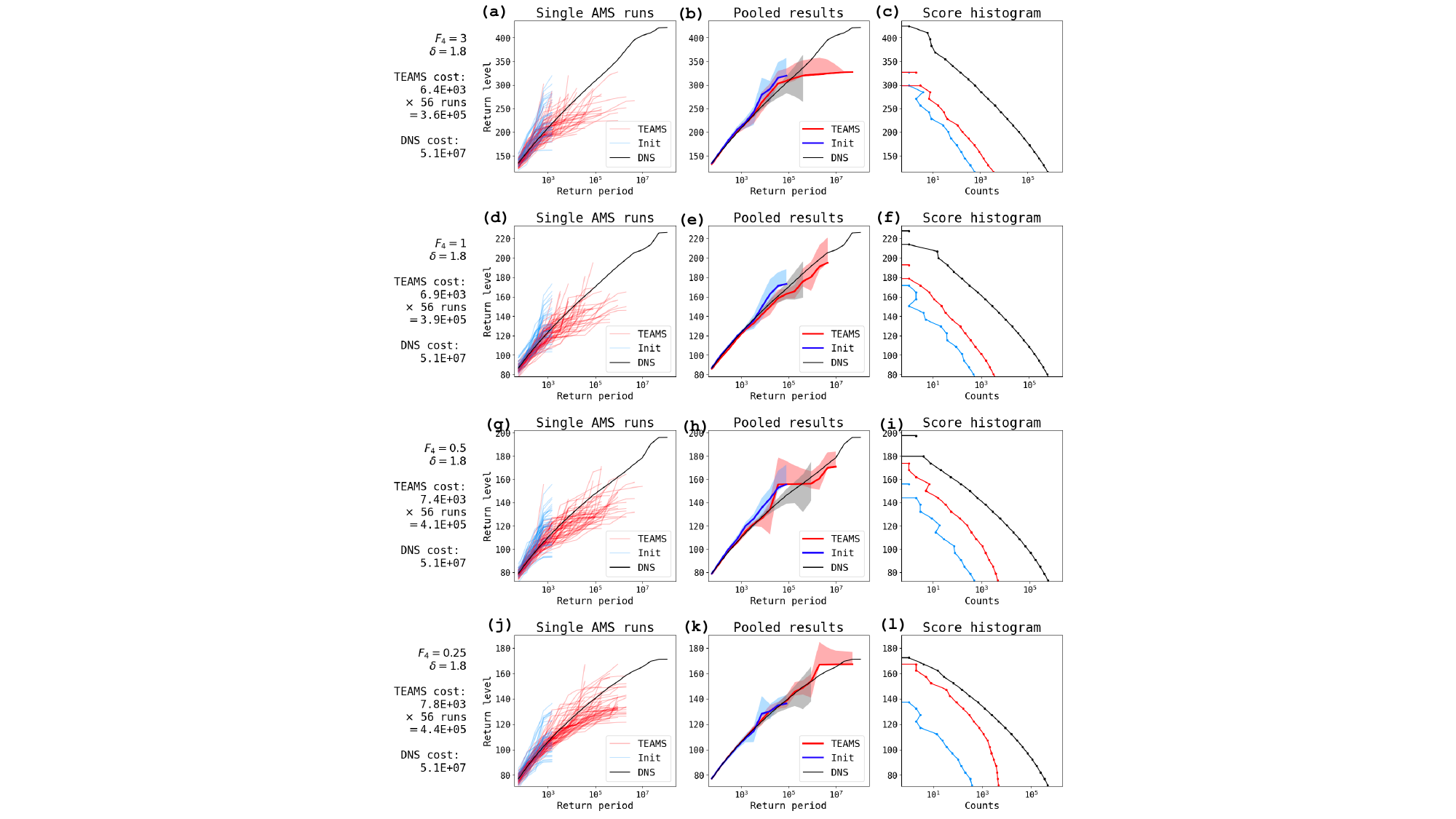"}
    \caption{TEAMS algorithm performance at all four noise levels with $\delta=1.8$}
    \label{fig:delta1p8}
\end{figure}

\begin{figure}
    \includegraphics[width=0.8\linewidth,trim={8cm 0cm 8cm 0cm},clip]{"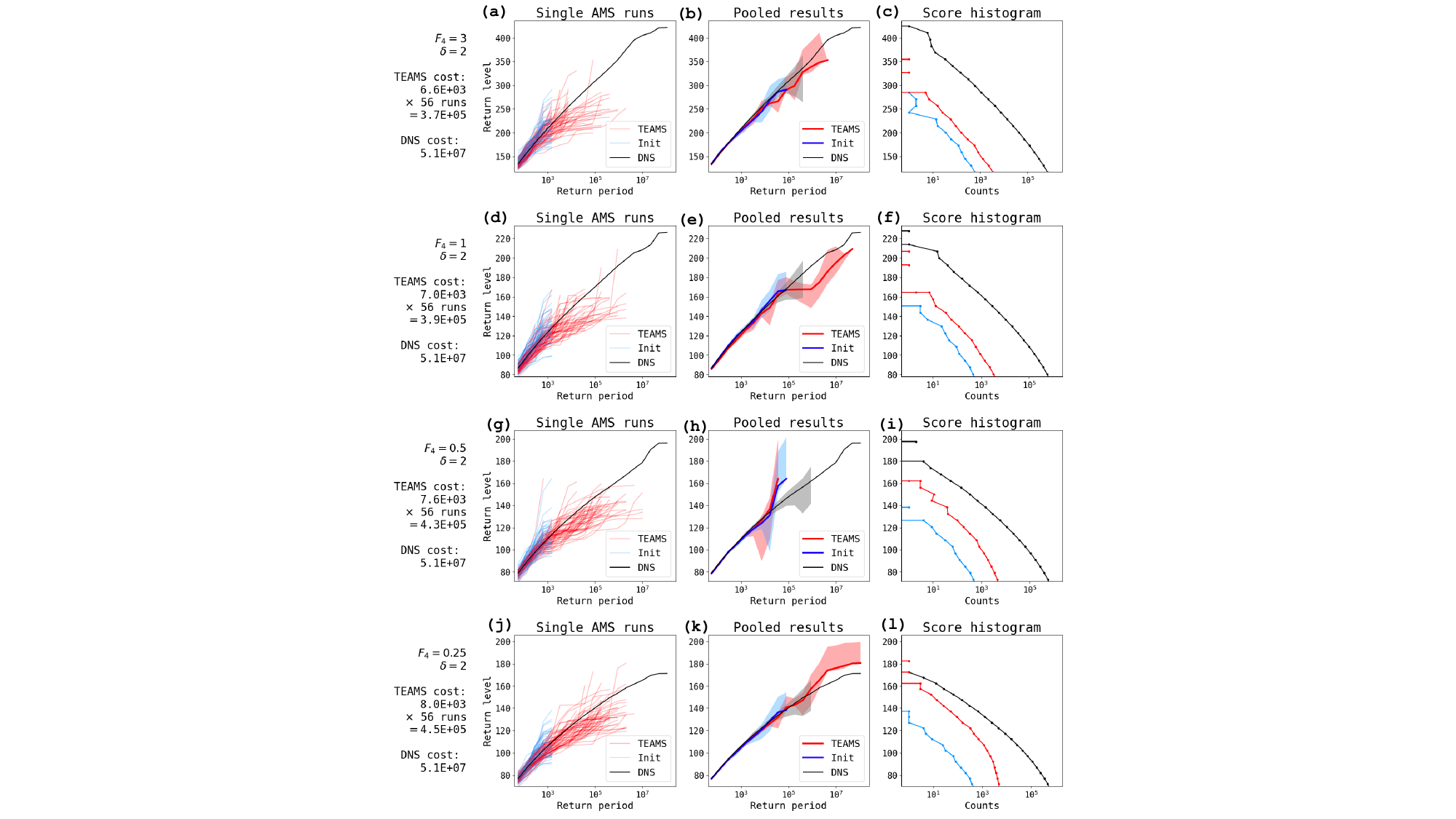"}
    \caption{TEAMS algorithm performance at all four noise levels with $\delta=2.0$}
    \label{fig:delta2p0}
\end{figure}

\end{document}